\documentclass[11pt]{article}
\usepackage{wrapfig,epsfig}
\newcommand{\gray}{$\gamma$-ray}
\newcommand{\grays}{$\gamma$-rays}
\newcommand{\etal}{{\it et al.}}

\newcommand{\amsg}{AMS/$\gamma$}
\begin{document}

\renewcommand{\baselinestretch}{1.7}
\small\normalsize

\begin{center}
{\Large\bf The Gamma Ray Detection Capabilities of the Alpha
Magnetic Spectrometer}
\end{center}

\vspace{2ex}

\begin{center}
{\bf R. Battiston$^{1}$, M. Biasini$^{1}$, E. Fiandrini$^{1}$,\\
J. Petrakis$^{2}$, and M.H. Salamon$^{2}$\\}

\vspace{0.5ex}

{\footnotesize $^{1}$Sezione INFN and Dipartimento di Fisica, Universit\`{a} di Perugia\\
$^{2}$Department of Physics, University of Utah\\}
\end{center}

\vspace{2ex}

\begin{abstract}
The modeled performance of the Alpha Magnetic Spectrometer (AMS) as a high-energy
(0.3 to 100 GeV) gamma-ray detector is described, and 
its gamma-ray astrophysics objectives are discussed.
\end{abstract}

\vspace{1ex}

\noindent
{\bf PACS, Keywords:} 95.55.K, 98.54.C, 95.35.+d, 98.35.G; high-energy gamma rays,
blazars, neutralinos, gamma-ray background

\newpage

\section{Introduction}
\label{intro.sec}

Our knowledge of the \gray\ sky has increased dramatically during this
last decade, due principally to the \gray\ instruments on board the
Compton Gamma Ray Observatory (CGRO)\cite{kur97}: EGRET,
a spark chamber plus calorimeter instrument with \gray\ sensitivity in the
energy interval 30 MeV to 30 GeV; COMPTEL, a Compton
telescope in the interval 0.1 to 30 MeV; BATSE, an
omnidirectional x-ray and soft \gray\ ``burst'' detector consisting of 
large NaI scintillators sensitive to 30 keV to 2 MeV photons (with smaller
spectroscopic NaI crystals for measurements up to 110 MeV); and 
OSSE, consisting of Nai-CsI phoswiches detecting photons of
0.1 to 10 MeV.  Their observations
have revolutionized our understanding of such extragalactic phenomena
as blazars and gamma ray bursts (GRBs), as well those within our
own Galaxy, such as pulsars.

A salient characteristic of these
extragalactic sources is their extreme time variability.  In the case
of blazars, flare states are observed to occur in which the luminosities
vary by over an order of magnitude within time-scales as small as hours.
GRBs are singular, extragalactic events due either to the
the coalescence of massive compact objects or to the hypernovae of massive
early stars, whose
luminosities decay over periods of seconds to days depending upon the
wavelength observed.  For such objects, the importance of correlated,
{\it multiwavelength} observations is paramount.  By monitoring the 
relative time lags of blazar flare peaks at radio, optical, x-ray,
and \gray\ wavelengths, for example, the properties of the source regions
at the bases of AGN (active galactic nucleus) jets are constrained.
Gamma ray observations of blazars by EGRET have been particularly crucial,
as the luminosity of these objects peaks in the GeV-to-TeV part of the
electromagnetic spectrum.  The transient nature of these sources therefore 
argues for {\it continuous} coverage of the \gray\ sky.

Until recently, observations with CGRO and other space-based
and ground-based telescopes have provided coverage of these
and other sources up to the limiting sensitive energy of EGRET, approximately
30 GeV.  From there, a gap in our knowledge of the \gray\ sky spectrum
has existed up to 200-300 GeV, where ground-based \gray\ shower detectors
presently have their energy thresholds.  It is possible that within this
gap there are novel features in the \gray\ sky, such as a gamma-ray line
or continuum emission from postulated neutralino annihilation at the 
center of the Galaxy.  Future instruments with sensivity in this
unexplored region, such as  AGILE \cite{agi98}, recently approved by
the Italian Space Agency (ASI),  or  the
proposed GLAST \cite{eng97} payload, may uncover exciting new   phenomena.

At the present time, however, our view of the \gray\ sky has diminished.
With the effective turnoff of the EGRET \gray\ instrument on the
CGRO due to the nearly
complete consumption of its spark chamber gas, there will soon
be no operating instrument capable of observing the \gray\ sky in
the energy interval $\sim 10^{-1}$ to $\sim 10^{2}$ GeV.
Ground-based \gray\ detectors, based on the atmospheric Cerenkov
technique (ACT)\cite{hil90}, turn on at current energy thresholds 
of $\sim$200-500 GeV.  Within
the next several years
energy thresholds for some ACT observatories are expected to go to as low as
20 to 50 GeV, but lower
thresholds than these are unlikely to be achieved due to the
sizable effect of Earth's magnetic field on the \gray -induced
air showers, and the lower Cerenkov photon yield which must be
detected against the night sky background.  For \gray\ energies much lower
than $10^{2}$ GeV, then, spaced-based detectors are required.  
This observational gap for the energy window $\sim 10^{-1}$ to $10^{2}$
GeV will continue to exist for the next several years.
Eventually, this gap will be eliminated
with the launch of a next-generation \gray\ satellite, such as GLAST \cite{eng97},
but such a mission is unlikely to occur before the year 2005.  Also the
AGILE satellite \cite{agi98}, to be launched in 
2002, will have  a limited sensitivity above
50 GeV.

%\begin{wrapfigure}[19]{r}{8.5cm}
%\begin{figure}
%\begin{center}
%\mbox{\epsfig{file=angres99-nim.eps,width=7.7cm}}
%\caption{\small Angular resolution of \amsg\ as a function of primary \gray\ energy.}
%\label{angres.fig}
%\end{center}
%\end{wrapfigure}
%\end{figure}
In this paper we describe how the Alpha Magnetic Spectrometer (AMS)
can largely fill this gap by acting as a \gray\ detector with sensitivity in the
energy interval of 0.3 to 100 GeV during its 
three-year mission on board the
International Space Station Alpha (ISSA) from 2003 to 2006.
AMS, described in detail elsewhere, has as its primary mission
the search for cosmic ray antinuclei, which are detected via the negative
curvature of their trajectories through a magnetic spectrometer 
(see Sect. \ref{ams.sec}).

It is of interest, however, to analyze if a large particle detector like
AMS, operated  on the Space Station for three years starting in 2003,  could
be used to detect high energy \grays\ . There are
at least   two ways  this could be done  without significantly affecting
the experiment sensitivity to antimatter: a) adding a light (e.g. $0.3 X_0$)
converter at the entrance of the magnetic spectrometer, either passive
(e.g., a high-Z thin plate) or active (e.g. a multilayer tracking detector), or
b) implementing an high granularity imaging shower detector at the bottom
of the experiment. Option (a) consists in  a minor modification to the
baseline instrument, while option (b), although would give a much higher
\gray\ detection capabilities,  requires the addition of a new detector.
In this paper we  present a study which has been perfomed on option (a). We
will show that in this  option, AMS can also detect \grays\ with
performance characteristics similar to EGRET in
the energy region of 0.3 to 20 GeV, and with significantly enhanced
capabilities between 20 and $\sim$ 100 GeV, a region which is not well
unexplored.  We refer to this modified instrument as ``\amsg''.
We show that \amsg\ can continue the valuable work of EGRET by providing
continued monitoring of extragalactic and galactic \gray\ sources and by
participating in multiwavelength observational campaigns.  In addition,
\amsg\ will have unprecedented sensitivity to the \gray\ sky between the
energies of 20 to $\sim 100$ GeV (albeit
at a level somewhat lower than required
for detection of known sources with power law spectra), so that
\amsg\ might provide us with unexpected discoveries in this region.

The next section gives a brief description of the baseline AMS instrument,
and the modifications required for adding \gray\ detection capability
in option (a).
Section 3 describes the performance characteristics of \amsg\ determined from
Monte Carlo analyses, and Section 4 uses these results to address the question
of what \gray\ astrophysics can be done with AMS.
The results obtained analyzing option (a)  give a hint of the physics
potential which
could be reached implementing option (b) for the flight of AMS on the Space
Station; option (b) would gain a factor of about $ 5$ in  \grays\
statistics over option (a).  Detailed analysis of option (b), however,  
wil be the subject of future work.

\section{The Alpha Magnetic Spectrometer}
\label{ams.sec}

\subsection{The Baseline Instrument}
\label{baseline.sec}

The Alpha Magnetic Spectrometer has been built by a large international
collaboration of high energy physics institutions from the U.S., Italy,
China, Finland, France, Germany, Taiwan, Russia, and Switzerland.
It recently
had a successful test flight on the Space Shuttle mission STS-91 in June, 1998
\cite{bat99},
where it was carried in the cargo bay and observed for several days in both
the zenith and nadir directions, the latter for measurement of albedo
cosmic ray backgrounds.  It is scheduled to be secured to an
external payload attachment point on the International Space Station Alpha
in 2003, where it will remain as a zenith-pointing instrument for 
a minimum of 3 years.

%\begin{figure}[ht]
%\begin{center}
%\mbox{\epsfig{file=amsearly.eps,width=8.5cm}}
%\caption{\small A cross section of the AMS instrument.}
%\label{ams-nim.fig}
%\end{center}
%\end{figure}

\begin{figure}[h]
\begin{center}
\mbox{\epsfig{file=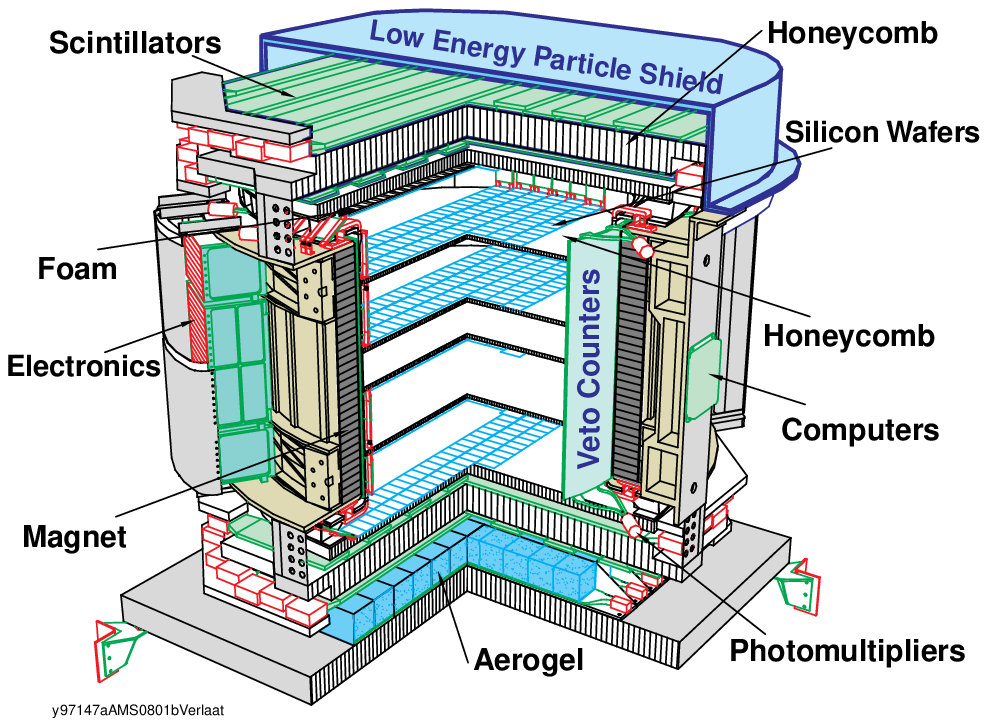,width=6.0in}}
\caption{\small A cross section of the AMS instrument.}
\label{ams-nim.fig}
\end{center}
\end{figure}

The basic design of AMS is shown in Figure \ref{ams-nim.fig}.  
The magnet spectrometer
consists of a permanent ring dipole magnet made of
very high grade Nd-B-Fe rare earth material whose magnetic energy product
and residual induction are respectively $(BH)_{\rm max}>50\times 10^{6}$
G-Oe and 14,500 G, yielding an analyzing power of $\left<BL^{2}\right>
=0.14$ T-m$^{2}$ with less than 2 tonnes of magnet mass.  

%\begin{figure}[htb]
%\begin{center}
%\mbox{\epsfig{file=y94725-nim.eps,width=7.7cm}}
%\caption{\small The momentum resolution of the AMS spectrometer for $e$, $\pi$,
%$K$, $p$, $^3$He, $^4$He, and $^12$C.}
%\label{y94725.fig}
%\end{center}
%\end{figure}

\begin{figure}[h]
\begin{center}
\mbox{\epsfig{file=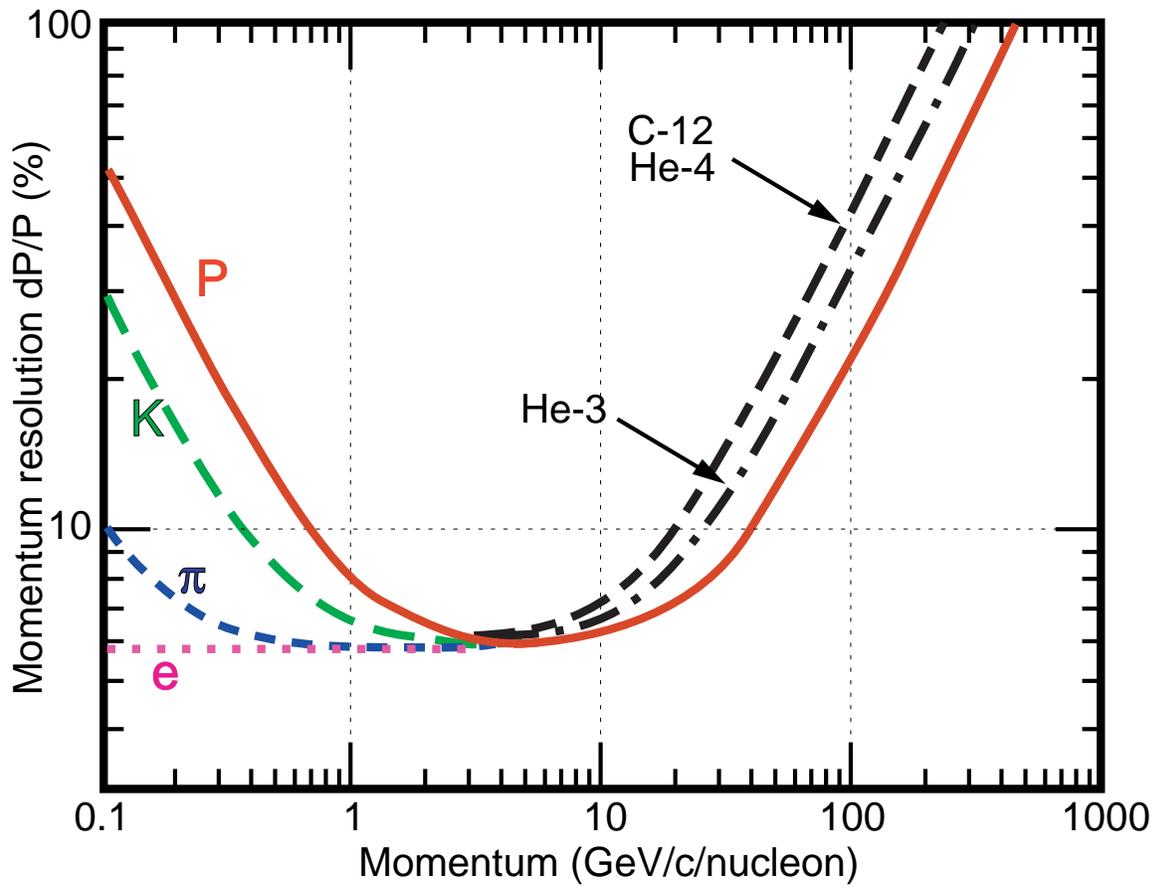,width=6.0in}}
\caption{\small The momentum resolution of the AMS spectrometer for $e$, $\pi$,
$K$, $p$, $^3$He, $^4$He, and $^{12}$C.}
\label{y94725.fig}
\end{center}
\end{figure}

Four silicon strip detector tracking planes are located
within the magnetic volume, with a fifth and sixth plane located
just above and below the magnet.
These are based on technology developed for
the Aleph and L3 vertex detectors at LEP \cite{bat95}, and consist
of $300 \mu$m thick, double-sided, high purity silicon wafers.  The total
area of this tracker is about an order of magnitude larger than the size
of previous microstrip silicon detectors.
The momentum resolution of the  silicon spectrometer is given in Figure \ref{y94725.fig}
at low rigidities (below 8 GV) the  resolution is dominated by multiple 
scattering  ($\Delta p/p\sim 7\% $),
while the maximum detectable rigidity ($\Delta p/p\sim 100\%$) is about $500$ GV. 
The parameters of the silicon spectrometer are given in Table 1.
In addition to measuring particle rigidity, the silicon planes will provide six
independent measurements of $dE/dx$ for charge determination. 

\begin{table}[t]
\caption{\small AMS silicon tracker parameters (in parenthesis the values used
on  the precursor flight).}
\begin{center}
\begin{tabular}{|c|c|} \hline
Number of planes & $6$  \\
Accuracy (bending plane) & $10\  \mu m$  \\
Accuracy (non bending plane) & $30\  \mu m$  \\
Number of channels & 163936   \\
                               & (58368)  \\
Power consumption & $400 \ W$ \\
                             & ($180 \ W$) \\
Weight                   & $130 \  kg$   \\
Silicon Area (double sided) & $5.4 \ m^2$   \\
         & ($2.4 \ m^2$)   \\ \hline
\end{tabular}
\end{center}
\end{table}

%%\begin{wrapfigure}{r}{8.5cm}
%\begin{figure}
%\begin{center}
%\mbox{\epsfig{file=area99-nim.eps,width=7.7cm}}
%\caption{\small (a) Histogram of detected \grays\ as a function of incident zenith angle
%$\theta$ for 10 GeV \grays.
%(b) Effective \gray\ detection area $A(E,\theta)$ versus zenith angle at 10 GeV.}
%\label{area.fig}
%\end{center}
%%\end{wrapfigure}
%\end{figure}
Four time-of-flight (ToF) scintillator planes (two above and two below the
magnet volume) 
measure the particle  velocity with a resolution of
120 ps over a distance of 1.4 m.  
The ToF scintillators also  measure $dE/dx$, allowing a
multiple determination of the absolute value of the particle charge.

A solid state Cerenkov detector below the magnet provides an
independent velocity measurement.
In addition, a scintillator anticounter system
is located within the inner magnet wall, extending to the ToF scintillators.

The performance of this instrument as a nuclear cosmic ray detector, and its
sensitivity to an antinucleus cosmic ray component, is discussed
elsewhere\cite{ahl94}.

We should also note that by the time AMS is attached to the International
Space Station it may have undergone significant changes from the baseline
design considered here.  In particular, the permanent magnet may be replaced
by a superconducting magnet, which would considerably improve
AMS's \gray\ detection performance (although it would increase the
\gray\ energy threshold as well; see \ref{energyres.sec}).  Since the
detector is still undergoing significant design changes from the baseline
instrument flown on the Space Shuttle, we have chosen to fix on that design
which currently exists as integrated hardware.  The addition of other
components, such as a transition radiation detector, a segmented calorimeter,
a solid \v{C}erenkov radiator, and/or a superconducting magnet will only
improve the performance of \amsg\ described in this paper.

\subsection{Modifications of AMS for \gray\ Detection}
\label{amsmod.sec}

In option (a), the conversion of AMS to \amsg\ requires the addition of 
two hardware components: (1) A high-$Z$ converter plate 
({\it e.g.} tungsten) of 0.3 radiation lengths ($x=0.3X_{0}$), which
converts \grays\ into electron-positron pairs.  The momenta
of each electron and positron are measured in the spectrometer,
and their weighted sums
give the primary photon's energy and incidence direction. (2) A stellar attitude
sensor gives the angular orientation of AMS with respect to the celestial
sphere to an accuracy of better than $1.5\times 10^{-4}$ radians.
The stellar attitude sensor is required because the Space Station cannot provide
attitude information to better than $\sim2^{\circ}$, while the AMS/$\gamma$
angular resolution approaches $1^{\prime}$ for $E\gamma > 100$ GeV.  Commercial units,
such as Ball Aerospace Systems Division's CT-633 Stellar Attitude Sensor, provide
angular resolutions to $0.5^{\prime}$ with very modest weight and power consumption
(2.5 kg and $<$ 10 W).

Determination of the converter thickness and position is based on the
following considerations:\\
(1) The \gray\ converter must not degrade the sensitivity of AMS for
antimatter detection.  Therefore the converter mass cannot be distributed
uniformly throughout the tracking volume, which would otherwise provide optimal
angular and energy resolution.  Instead, the converter must lie as a single plate
just above the second TOF scintillator, S2.\\
(2) The loss of nuclear antimatter candidates to nuclear interactions in the
converter plate must be insignificant:  At $x=0.3X_{0}$, this amounts to only 
$\sim2$\% of cosmic-ray $^{4}$He being lost.\\
(3) The sensitivity for detection of astrophysical point sources of gamma rays
is nearly independent of $x$ for $x\ll X_{0}$ for background-limited (as opposed
to statistics-limited) detection.
This sensitivity degrades as $x/X_{0}\rightarrow 1$.\\
(4) The aperture for rare events and for measurement of the diffuse
gamma ray background increases with $x$.\\
(5) The fractional energy loss of the electron pair due to bremsstrahlung within
the converter is $\propto x/X_{0}$, degrading energy resolution as $x$ increases.\\
(6) Can the presence of the converter plate create background that will degrade
the cosmic ray $\overline{\rm He}/{\rm He}$ target sensitivity of $10^{-9}$?
By requiring clean tracks in the spectrometer volume we do not expect
converter-generated contamination, as the converter lies above the magnet volume.
This is corroborated both by the SHuttle flight data (where an 
$\overline{\rm He}/{\rm He}$ of $\sim 10^{-6}$ was obtained \cite{bat99})
and by Monte Carlo analyses.  This question is also being examined using
helium and carbon data from GSI at Darmstadt.\\

For this anaysis we choose the value $x=0.3X_{0}$, 
for which point source sensitivity is still optimal,
the energy resolution is acceptable, and nuclear interaction losses are
negligible.

\section{The Performance of AMS/$\gamma$}
\label{performance.sec}

\subsection{Monte Carlo Calculations}
\label{mc.sec}

A full-instrument GEANT Monte Carlo code was run to determine the 
performance of \amsg. Gamma rays with fixed energies, ranging 
from 0.3 to 100 GeV, were
thrown isotropically at the detector over an opening angle of 
$50^{\circ}$.  

%%\begin{wrapfigure}{r}{8.5cm}
%\begin{figure}
%\begin{center}
%\mbox{\epsfig{file=aperture99-nim.eps,width=7.7cm}}
%\caption{\small \amsg\ aperture as a function of \gray\ energy.  The statistical errors
%are comparable in size to the circular data points.}
%\label{aperture.fig}
%\end{center}
%%\end{wrapfigure}
%\end{figure}
For those \grays\ which converted in the detector (most
within the converter plate) the charged particles were tracked through the
inhomogeneous field of the ring dipole permanent magnet.  The inhomogeneous
field used was that measured for the AMS magnet that was recently flown
on the Space Shuttle (flight STS-91), with a field strength increased by
7\% to match the expected field strength of a second magnet 
which might be constructed for the Space Station mission.

All the
physical processes for electrons and \grays\ were ``on'' in the GEANT
code.  Bremsstrahlung photons of energies $<20$ MeV were not followed,
although all bremsstrahlung energy losses were included.  The hit positions in
the silicon tracker planes were assigned gaussian errors with $\sigma=15\mu$m
and $30\mu$m for the bending and nonbending planes, respectively.  (Test
beam measurements give a bending plane Si tracker resolution below
10 $\mu$m \cite{amb98}.)
Reconstruction of the particle trajectories were performed
with a quintic spline fit routine \cite{win74}
to determine their momenta and charge sign.  Assignment of a hit position
within the 6-plane Si tracker to either
the converted electron or positron was done for all possible combinations
(64 in all) of assignments; that 
combination whose trajectory reconstruction $\chi^{2}$ was minimum
was selected as the event.
All secondary charged particles
were tracked in the detector volume until they stopped or exited from
the detector mother volume.  

%\begin{wrapfigure}{r}{8.5cm}
%\begin{
%\begin{center}
%\mbox{\epsfig{file=hammer-nim.eps,width=7.7cm}}
%\caption{\small A Hammer-Aitoff projection in galactic coordinates 
%of the AMS area$\times$time factor at 50 GeV for
%one year live-time operation of AMS. Contours are shown in units of
%$10^{9}$ cm$^{2}\cdot$s.  The Galactic Center (GC) is at galactic longitude
%$l=0$ and galactic latitude $b=0$.  The locations of the
%north and south celestial poles (NCP, SCP) are
%also indicated, as are those of the Crab nebula and several prominent blazars.}
%\label{hammer.fig}
%\end{center}
%\end{wrapfigure}
% \end{figure}

The \gray\ candidate
events were those that passed the hardware trigger of
$\overline{S1}\cdot S2 \cdot S3 \cdot S4$, where $S1$ to $S4$ are the
ToF counters (the converter plate lies between $S1$ and $S2$).  If reconstruction
of particle trajectories occurred successfully, further cuts were
placed on these candidate events before acceptance as a \gray\ event:\\
\ \ 1) the reconstructed \gray\ zenith angle $\theta_{\gamma}$
must be $\leq 50^{\circ}$;\\
\ \ 2) the reconstructed \gray\ energy $E_{\gamma}$ must be
$\geq 0.15$ GeV;\\
\ \ 3) there are at least two reconstructed tracks;\\
\ \ 4) the reconstructed energy of the most energetic particle be $\geq 0.15$ GeV;\\
\ \ 5) the number of anticoincidence panels hit (there are 16 such panels 
along the interior wall of the magnet) be $<4$.

Primary \gray\ energy and incidence direction were determined by adding the fitted
momenta vectors of all secondaries evaluated at the converter plate to obtain
the primary momentum vector.  (More sophisticated weighting methods were found
to have negligible effect on \gray\ angular resolution.)
The results of the Monte Carlo runs are discussed in the subsections below.

\subsection{Point Source Sensitivity}
\label{sensitivity.sec}

Astrophysical sources of \grays\ fall into two categories: point sources, such
as blazars and GRBs, and diffuse sources, such as the cosmic isotropic \gray\ 
background and possibly \gray\ emission from neutralino annihilation near the
center of our Galaxy.   Some galactic sources fall into an intermediate
category of ``extended'' sources, such as pulsar nebula and supernova remnants,
where the source angular size may be larger than the point spread function of the
\gray\ detector.

For point sources, a detector's key figure of merit is its {\it point source
sensitivity}, defined as the minimum source flux required to achieve a specified
level of detection significance.  The significance $S$ of a detection is
given schematically by
\begin{equation}
S(>E_{\rm t})\sim \frac{N(>E_{\rm t})}{B(>E_{\rm t})^{1/2}},
\label{eq1}
\end{equation}
where $N(>E_{\rm t})$ and $B(>E_{\rm t})$ are respectively the total number
of detected photons from the source and the number of background photons falling
within the source area above an energy $E_{\rm t}$.  $E_{\rm t}$ is usually, but not
always, the instrument's threshold energy.  $N$ and $B$ are a function of the
effective detection area $A(E)$ of the instrument, its angular resolution
expressed as a solid angle $\Omega(E)$, the viewing time $t$, and the differential
source and background spectra $dN/dE$ and $dB/dE$:
\begin{equation}
N(>E_{\rm t})=\int_{E_{\rm t}}^{\infty}\frac{dN}{dE}A(E)t\,dE,
\label{eq2}
\end{equation}
and
\begin{equation}
B(>E_{\rm t})=\int_{E_{\rm t}}^{\infty}\frac{dB}{dE}A(E)\Omega(E)t\,dE.
\label{eq3}
\end{equation}

All extragalactic objects (blazars) seen by EGRET are well-represented by power law
spectra
\begin{equation}
\frac{dN}{dE}=n_{0}\left( \frac{E}{\rm 1 GeV} \right)^{-\gamma}
({\rm cm}^{2}\cdot{\rm s}\cdot{\rm GeV})^{-1},
\label{eq4}
\end{equation}
where $\gamma$ is the differential spectral index of the 
source, and $n_{0}$ is the differential
flux at 1 GeV.  We define our point source sensitivity as that value
of $n_{0}$ which gives a $5\sigma$ signal integrated over the period of one year
of operation of AMS.  It is evident that the point source sensitivity is
a function of the source's spectral index; for these calculations we have assumed
a source differential spectral index of $\gamma=2$, which is close to the
mean of the source spectral indices of blazars observed by EGRET 
\cite{muk97}.

The background flux $dB/dE$ contains components from the isotropic extragalactic
\gray\ background radiation 
and from the galactic diffuse radiation
(the latter due to the decay of $\pi^{0}$s produced in collisions of
cosmic rays with the interstellar medium).  The former has recently been
determined by EGRET to be \cite{sre98}
\begin{equation}
\frac{dB_{\rm extragal}}{dE}=(7.32\pm 0.34)\times 10^{-6}
\left(\frac{E}{0.451 {\rm \ GeV}}\right)^{2.10\pm 0.03}
({\rm cm}^{2}\cdot{\rm s}\cdot{\rm sr}\cdot{\rm GeV})^{-1}.
\label{egrb.eq}
\end{equation}
The galactic diffuse radiation is a strong function of position in the sky,
and been measured and modelled by the EGRET group \cite{hun97}.  We have used
here their data to generate background spectra $\frac{dB}{dE}(E,\delta,\alpha)$
over $3^{\circ}$ bins in declination $\delta$ and right ascension $\alpha$.

\begin{figure}[h]
\begin{center}
\mbox{\epsfig{file=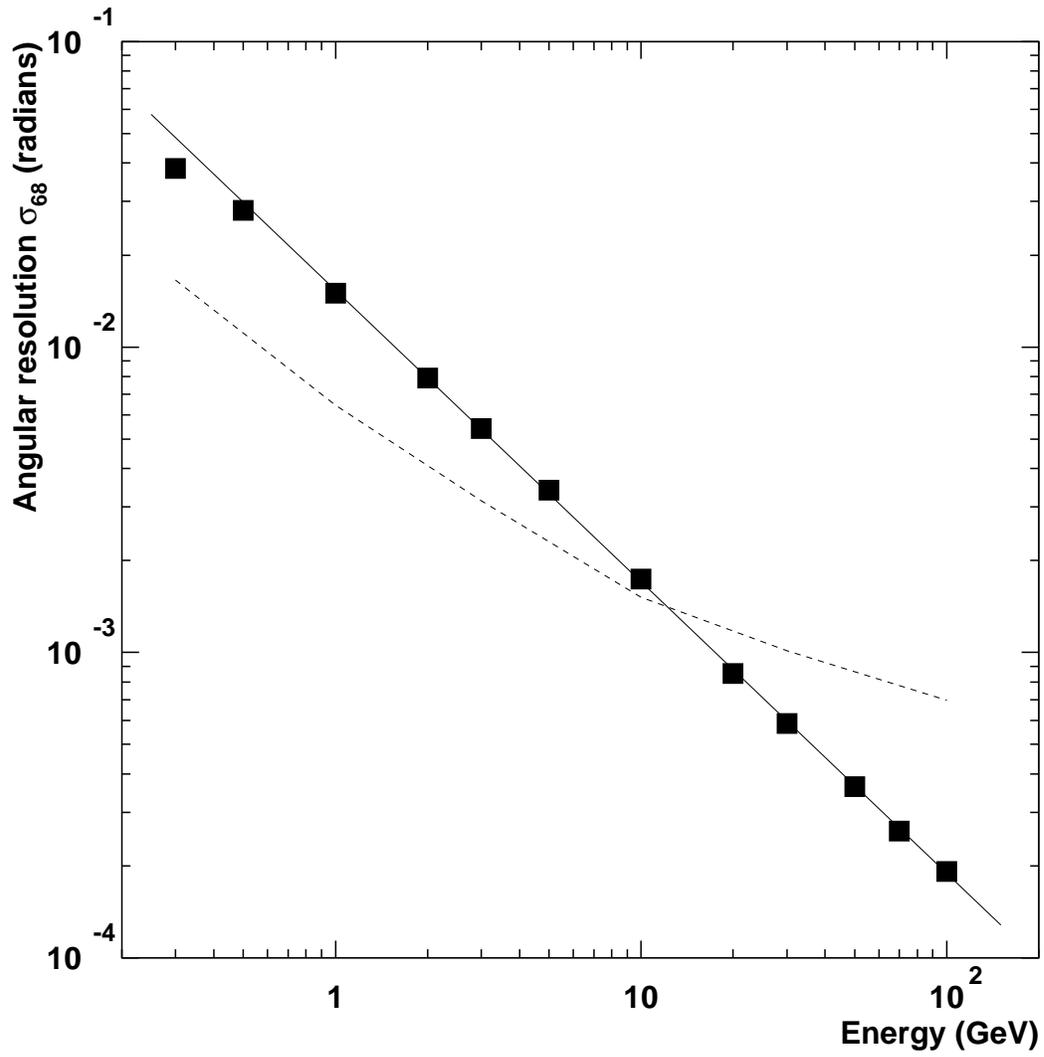,width=6.0in}}
\caption{\small Angular resolution of \amsg\ as a function of primary \gray\ energy
(filled squares), and that of GLAST \cite{blo96} (dashed line)
in the energy interval 0.3 to 100
GeV.}
\label{angres.fig}
\end{center}
\end{figure}

The solid angle $\Omega(E)$ over which the background must be integrated when
viewing a source is given by $\Omega(E)=\pi\sigma^{2}_{68}(E)$, where
$\sigma_{68}$, following EGRET's usage, is defined as the angular radius within
which 68\% of the source photons fall.  The converter thickness of 0.3 $X_{0}$
largely dominates all multiple scattering effects,
so that the angular and energy resolution of reconstructed primary photons
will be completely dominated by multiple Coulomb scattering (MCS) and bremsstrahlung energy
losses of the electrons within the converter plate.  Figure
\ref{angres.fig} confirms that the Monte Carlo result for
\amsg 's angular resolution is very close to that estimated from MCS
within the converter plate:
\begin{equation}
\sigma_{68}^{\rm AMS}(E)=0.88^{\circ}\left(\frac{E}{1 {\rm \ GeV}}\right)^{-0.956},
\label{eq6}
\end{equation}
which is to be compared to EGRET's angular resolution \cite{tho93} of
\begin{equation}
\sigma_{68}^{\rm EGRET}(E)=1.71^{\circ}\left(\frac{E}{1 {\rm \ GeV}}\right)^{-0.534}.
\label{eq7}
\end{equation}

\begin{figure}[h]
\begin{center}
\mbox{\epsfig{file=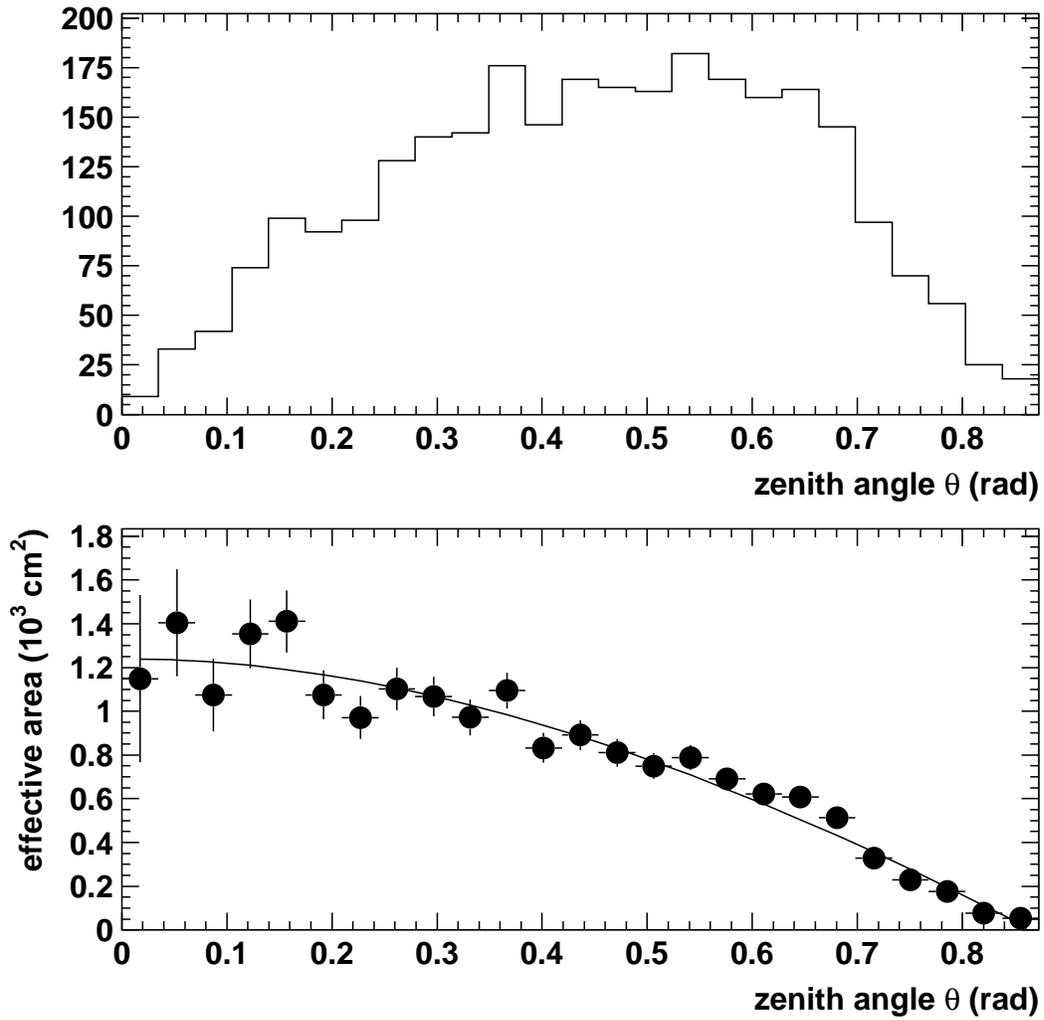,width=6.0in}}
\caption{\small (a) Histogram of detected \grays\ as a function of incident zenith angle
$\theta$ for 10 GeV \grays.
(b) Effective \gray\ detection area $A(E,\theta)$ versus zenith angle at 10 GeV.}
\label{area.fig}
\end{center}
\end{figure}

\begin{figure}[h]
\begin{center}
\mbox{\epsfig{file=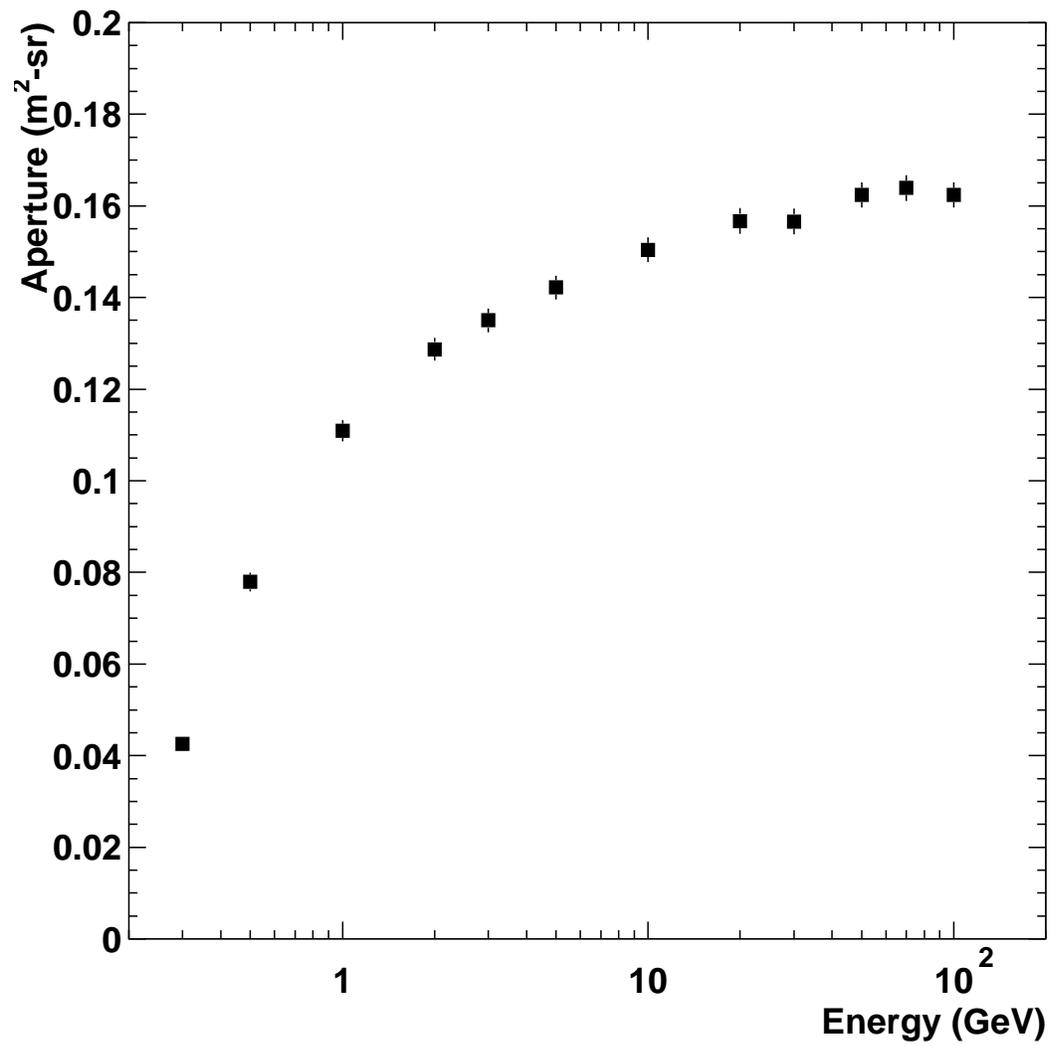,width=6.0in}}
\caption{\small \amsg\ aperture as a function of \gray\ energy.  The statistical errors
are comparable in size to the circular data points.}
\label{aperture.fig}
\end{center}
\end{figure}

%%\begin{wrapfigure}{r}{8.5cm}
%\begin{figure}
%\begin{center}
%\mbox{\epsfig{file=areatime99-nim.eps,width=7.7cm}}
%\caption{\small Accumulated area*time product over one year of AMS operation as
%a function of declination; each curve is marked by the \gray\ energy (in GeV).
%The declination of the Galactic Center (GC) is also indicated as a dashed 
%vertical line.}
%\label{areatime.fig}
%\end{center}
%%\end{wrapfigure}
%\end{figure}
The effective area $A(E)$ of the instrument is also a function of the zenith
angle $\theta$ of the incident photon, as shown in Figure \ref{area.fig}
for 3 GeV photons.  Although the $\theta$ dependence of the conversion
probability cancels that of the projected area,
the requirement that $S3$ and $S4$ ToF scintillators
be hit causes the effective area to drop to zero for $\theta>50^{\circ}$.
An integration of effective area $A(E,\theta)$ over solid angle gives 
instrument aperture, shown in Figure \ref{aperture.fig}.
Below a \gray\ energy of 0.3 GeV, the converted electrons begin to have too small a radius
of curvature to escape from the magnet volume, and detection efficiency plummets.
Above 100 GeV 
the converted electron and positron often do not spatially diverge beyond the
two-hit resolution distance of the silicon trackers, causing significant deterioration
in \gray\ energy resolution.  These
then define the limits of the energy window for \amsg.

\begin{figure}[h]
\begin{center}
\mbox{\epsfig{file=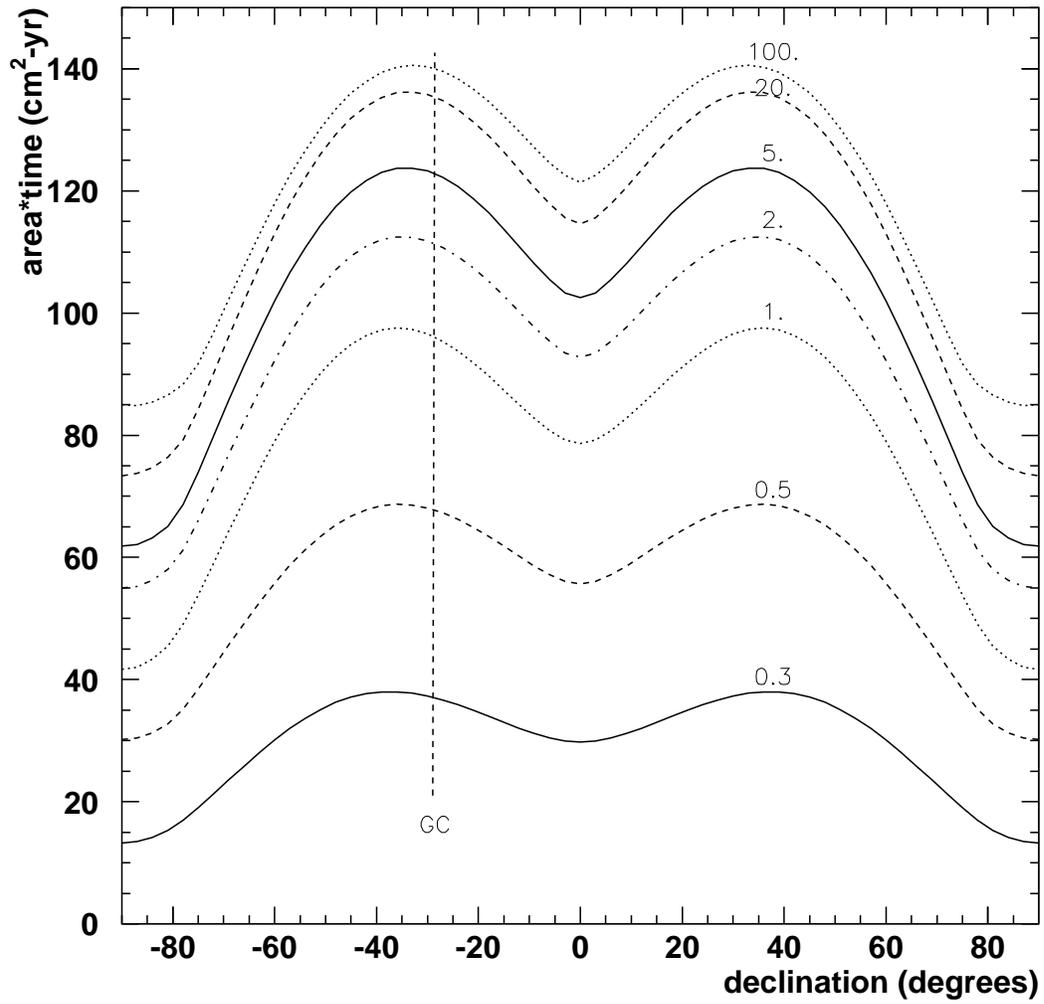,width=6.0in}}
\caption{\small Accumulated area*time product over one year of AMS operation as
a function of declination; each curve is marked by the \gray\ energy (in GeV).
The declination of the Galactic Center (GC) is also indicated as a dashed 
vertical line.}
\label{areatime.fig}
\end{center}
\end{figure}

Because AMS will be rigidly attached to the ISSA which itself will be in 
51.6$^{\circ}$ inclination orbit, AMS will not spend equal amounts of time
viewing all directions in the celestial sphere. However, \amsg\ 
has finite detection area out to $\theta=50^{\circ}$, so that \amsg\ does view the
entire sky.  Full sky coverage is obtained roughly every three months due to
the precession of the orbital plane of the Space Station about Earth's pole
\cite{jsc95}.  Because of this precession, the viewing area$\times$time product over
the sky is only a function of $E$ and declination $\delta$, and not of 
right ascension $\alpha$.  By following the orbit of AMS over one year, at
each time element determining the $A\Delta t$ increment for each position in
the sky viewable by AMS at that time, a total $At(E,\delta)$ area$\times$time
product is obtained, and is shown in Figure \ref{areatime.fig}.  

\begin{figure}[h]
\begin{center}
\mbox{\epsfig{file=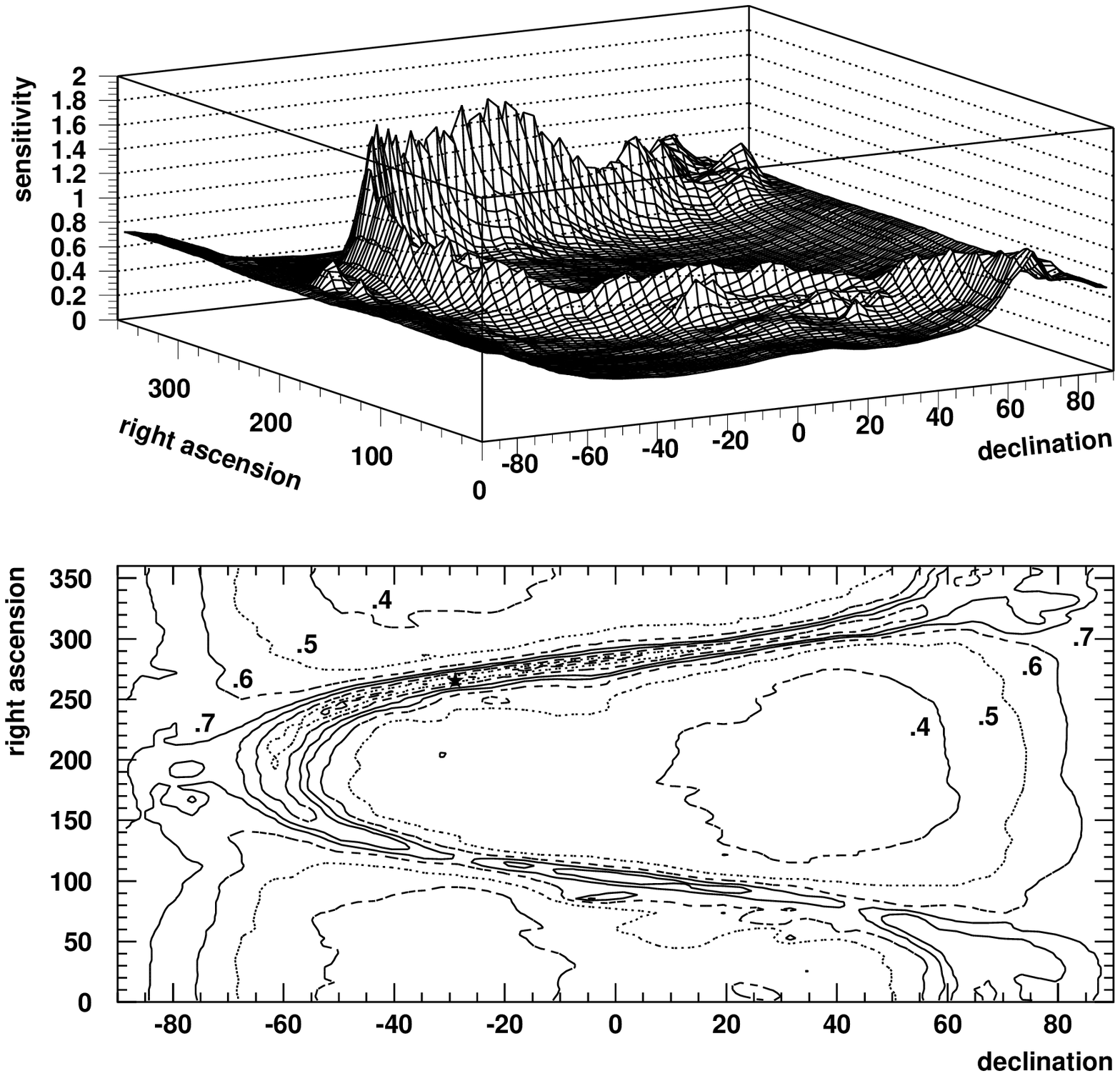,width=6.0in}}
\caption{\small A 3-D plot and 2-D contour plot of \amsg 's point source
sensitivity $n_{0}$ versus celestial coordinates.  Recall that $n_{0}$ is
the minimum amplitude of the source's differential flux at 1 GeV required for
a $5\sigma$ significance detection.  The units for $n_{0}$ in the 3-D plot are
$10^{-8}$ cm$^{-2}\cdot$s$^{-1}\cdot$GeV$^{-1}$.  Large photon fluxes from the diffuse
galactic background
are responsible for the deterioration of sensitivity near the Galactic plane.}
\label{sens.fig}
\end{center}
\end{figure}

With the above results and Eqs. \ref{eq2}-\ref{eq6}, \amsg 's 
$5\sigma$ point source sensitivity can be obtained.  The significance is calculated
using the likelihood ratio of Li and Ma \cite{li83},
\begin{equation}
\label{eq8}
S=\sqrt(2)\left\{N_{\rm on}\ln\left[\frac{1+\alpha}{\alpha}\left(
\frac{N_{\rm on}}{N_{\rm on}+N_{\rm off}}\right)\right]+
N_{\rm off}\ln\left[(1+\alpha)\left(\frac{N_{\rm off}}{N_{\rm on}+N_{\rm off}}
\right)\right]\right\}^{1/2},
\end{equation}
where $\alpha\equiv t_{\rm on}/t_{\rm off}$, $t_{\rm on}$ being the time spent
viewing the source, and $t_{\rm off}$ being the time spent viewing off-source,
and $N_{\rm on}$ and $N_{\rm off}$ are the total integrated on-source (signal plus
background) and
off-source (background) photon counts.  
Since at 1 GeV roughly $10^{-3}$ of the instrument's
field-of-view (FOV) is contained in the $\sigma_{68}$ error circle, we take
$\alpha=10^{-3}$.  For any given position in the sky, that computed value 
of $n_{0}$ which gives $S=5$ at 1 GeV is the point source sensitivity for that
sky position.  Figure \ref{sens.fig} shows the one-year point source
sensitivity for a lower cutoff energy $E_{\rm t}$ of 1 GeV.

%%\begin{wrapfigure}{r}{8.5cm}
%\begin{figure}
%\begin{center}
%\mbox{\epsfig{file=sens99-nim.eps,width=7.7cm}}
%\caption{\small A 3-D plot and 2-D contour plot of \amsg 's point source
%sensitivity $n_{0}$ versus celestial coordinates.  Recall that $n_{0}$ is
%the minimum amplitude of the source's differential flux at 1 GeV required for
%a $5\sigma$ significance detection.  The units for $n_{0}$ in the 3-D plot are
%$10^{-8}$ cm$^{-2}\cdot$s$^{-1}\cdot$GeV$^{-1}$.  Large photon fluxes from the diffuse
%galactic background
%are responsible for the deterioration of sensitivity near the Galactic plane.}
%\label{sens.fig}
%\end{center}
%%\end{wrapfigure}
%\end{figure}
One cannot directly compare the point source sensitivity of \amsg\ 
to that of EGRET, since AMS is not a pointable instrument.
EGRET achieves a flux sensitivity of $I_{\rm min}(>0.1 {\rm \ GeV})
\approx 10^{-7} {\rm cm}^{-2}{\rm s}^{-1}$ with a 2-week viewing period.
Since it took EGRET one year to map the full sky, where each sky segment
was viewed for roughly 2 weeks, we can compare EGRET's sensitivity with that 
achieved by \amsg\ after one year of operation.
By assuming a source differential spectrum of $E^{-2}$, we can convert from
EGRET's definition of sensitivity (in terms of integral flux above 0.1 GeV)
to that of \amsg\ (a differential flux above 1 GeV).  In our units the
EGRET $5\sigma$ flux sensitivity is $1\times 10^{-8}$.
Over most of the sky, \amsg 's 
mean sensitivity $\left<n_{0}\right>$ is estimated to
be about a factor of 2 lower than that of EGRET.

\subsection{Energy Resolution}
\label{energyres.sec}

\begin{figure}[h]
\begin{center}
\mbox{\epsfig{file=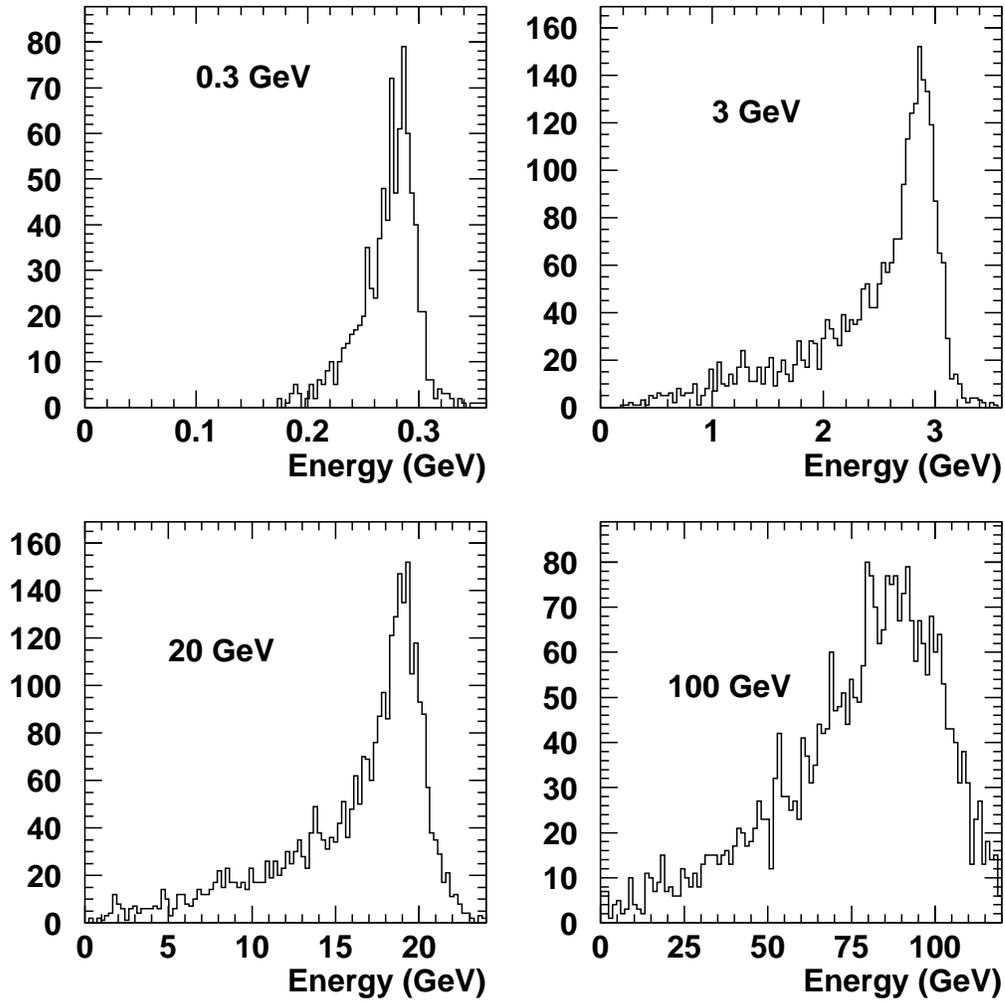,width=6.0in}}
\caption{\small The reconstructed energy distributions for four different primary
\gray\ energies, 0.3, 3, 20, and 100 GeV.}
\label{energy.fig}
\end{center}
\end{figure}

\begin{figure}[h]
\begin{center}
\mbox{\epsfig{file=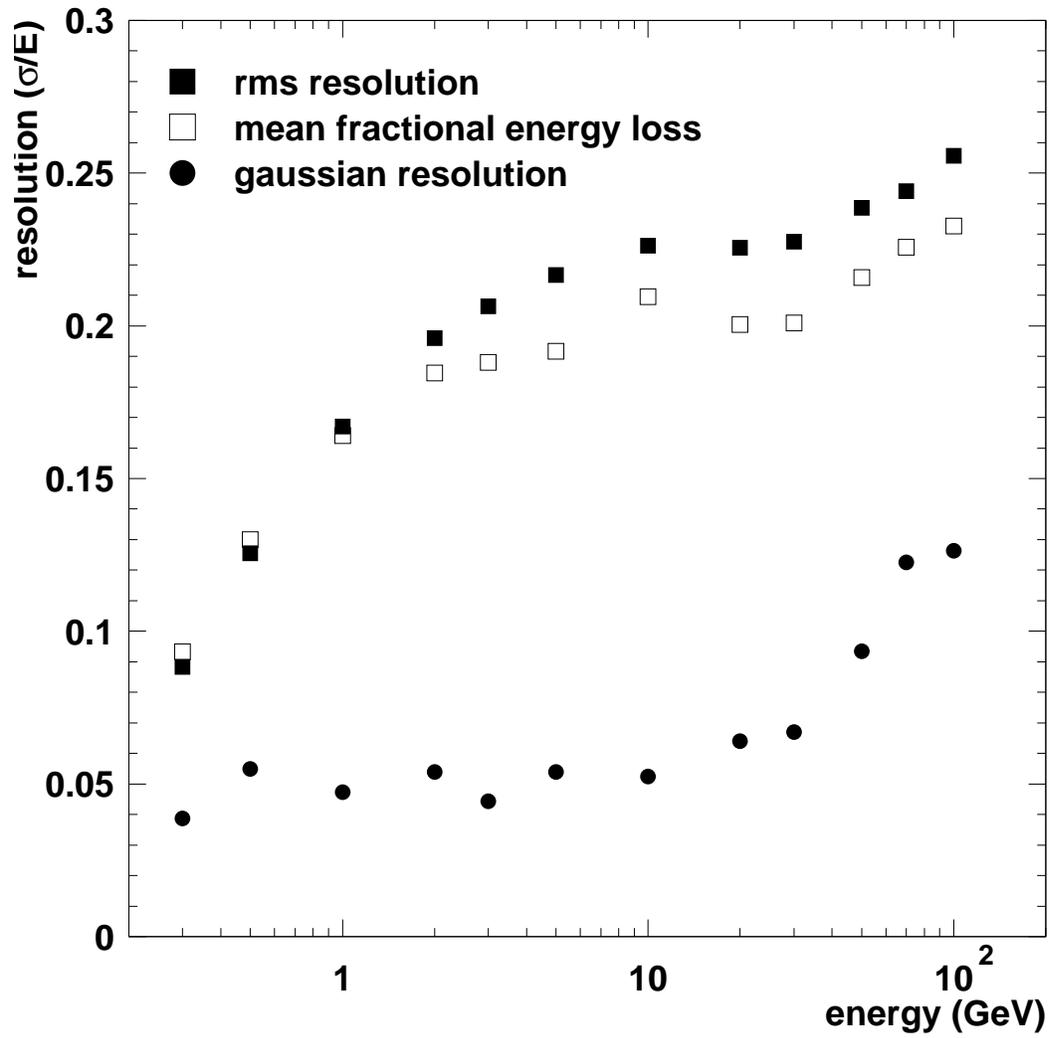,width=6.0in}}
\caption{\small The mean fractional energy loss (open square), rms energy resolution
(filled square), and width of the ``gaussian'' peak of the distribution (filled
circle) as a function of primary \gray\ energy.}
\label{energyres.fig}
\end{center}
\end{figure}

The \gray\ energy resolution is dominated by bremsstrahlung losses in the
converter plate, which have an average value of 
$\left<\frac{\Delta E}{E}\right>\approx \frac{1}{2}\left(
\frac{7\tau}{9X_{0}\cos\theta}\right)\approx 0.16/\cos\theta$ for
a converter thickness $\tau=0.3 X_{0}$.  
%%\begin{wrapfigure}{r}{8.5cm}
%\begin{figure}
%\begin{center}
%\mbox{\epsfig{file=energy99-nim.eps,width=7.7cm}}
%\caption{\small The reconstructed energy distributions for four different primary
%\gray\ energies, 0.3, 3, 20, and 100 GeV.}
%\label{energy.fig}
%\end{center}
%%\end{wrapfigure}
%\end{figure}
In the reconstructed \gray\ energy
distributions for primary energies of 0.3, 3, 20, and 100 GeV, shown in
Figure \ref{energy.fig},
one sees the effect of bremsstrahlung losses in the converter plate in the
large low-energy tails of the distributions.  The nearly identical appearances of these
distributions is a consequence of the fact that 
the probability for the electron to radiate
a fraction $\nu$ of its energy as a bremsstrahlung photon is 
solely a function of $\nu$.  There are two noticeable deviations from this
universality in the figure: The cutoff of events below 0.1 GeV is due
to the gyroradii of the electrons becoming smaller than the magnet field
volume dimensions; the broadening of the distribution at 100 GeV is
due to the fact that trajectory reconstruction errors are beginning to dominate
bremsstrahlung losses in the energy determination.  Figure \ref{energyres.fig}
gives the mean fractional energy loss and the rms energy resolution as a function
of primary \gray\ energy, and shows the deviations from constant values below 1 GeV and
near 100 GeV.

%%\begin{wrapfigure}{r}{8.5cm}
%\begin{figure}
%\begin{center}
%\mbox{\epsfig{file=energyres99-nim.eps,width=7.7cm}}
%\caption{\small The mean fractional energy loss (open square), rms energy resolution
%(filled square), and width of the ``gaussian'' peak of the distribution (filled
%circle) as a function of primary \gray\ energy.}
%\label{energyres.fig}
%\end{center}
%%\end{wrapfigure}
%\end{figure}
That the reconstructed energy distributions have a nearly ``universal'' form
has important consequences 
for the determination of blazar energy spectra.  
If a source energy spectrum follows a
power-law energy distribution, $E^{-\gamma}$, and 
if the energy resolution function of the detector is of the form
$f(E^{\prime},E)=E^{-1}g(E^{\prime}/E)$, where $E$ is the true incident energy,
$E^{\prime}$ is the reconstructed energy, and $g(x)$ is an arbitrary
function of $x$, then the reconstructed
source spectrum is also a power-law spectrum {\it with the same spectral index}.
There will therefore be little bias
in our reconstructed spectral indices.  This is demonstrated in Section \ref{ib.sec}
(Figure \ref{blazaratt-nim.fig})
where the measured spectra of several blazars, a convolution of the primary 
\gray\ energy
spectrum $E^{-\gamma}$ and the energy resolution function $f(E^{\prime},E)$,
are seen to match the true spectra.

\subsection{Comparison of \amsg\ with EGRET}

The table below summarizes the performance characteristics of AMS as a \gray\ 
detector.  In particular, by its comparison to EGRET, one sees that
the two instruments perform similarly in many respects, one major difference being
the energy windows: AMS's energy window is shifted up by roughly one order of magnitude from
that of EGRET, thereby providing an improved view of the sky in the region
$E_{\gamma}\sim 100$ GeV.\\

\begin{tabular}{|l|c|c|} \hline
 & AMS & EGRET \\ \hline \hline
technique & magnetic spectrometer & spark chamber plus calorimeter \\ \hline
energy window (GeV) & 0.3 to 100. & 0.03 to 30. \\ \hline
peak effective area (cm$^{2}$) & 1300 & 1500 \\ \hline
angular resolution & $0.77^{\circ}(E/1 GeV)^{-0.96}$ & $1.71^{\circ}(E/1 GeV)^{-0.534}$ \\ \hline
half-area zenith angle & $\sim 30^{\circ}$ & $\sim 20^{\circ}$ \\ \hline
total viewing time (yr) & $\sim$3 & $\sim$2 \\ \hline
attitude capability & fixed & movable \\ \hline
flux sensitivity & $\sim0.5\times10^{-8}$ & $\sim1.0\times10^{-8}$ \\
(ph/cm$^{2}$-s-GeV at 1 GeV) & & \\ \hline
\end{tabular}
\vspace{1ex}

A few of the table entries require some amplification:  (1) The 30 GeV upper limit
listed for EGRET is due in part to the effects of electromagnetic backsplash from
their calorimeter into their hardwired anticoincidence scintillator shield that
surrounds the instrument.  Above $\sim$GeV energies this self-induced
veto reduces EGRET's aperture 
roughly as $E^{-0.5}$.  Although the EGRET collaboration
quotes 30 GeV as their maximum energy, in fact they have small but finite aperture
up to and beyond 100 GeV.  (2) The total viewing time for EGRET is listed as
2 years.  This takes into account the fact that, on average, 
during approximately one-third of the viewing time the pointing direction
is occulted by Earth.  (3)  Unlike
EGRET, \amsg\ is not a pointing instrument; its vertical axis is always
aligned with the zenith.  Therefore \amsg\ cannot point to targets of
opportunity.

Note that the point source sensitivity
is nearly the same for both EGRET and AMS,
with that of AMS being somewhat (factor $\sim$2) lower (given the caveat
of Section \ref{sensitivity.sec}).
This is because EGRET and AMS have 
similar effective detection areas and angular resolutions.
The fact that AMS's \gray\ energy threshold is an order of magnitude
larger than EGRET's, thereby implying an integrated point source flux about an order of
magnitude lower for AMS than for EGRET, is compensated for by the fact that AMS's angular
aperture (half-angle area of $30^{\circ}$) is larger than EGRET's ($\sim 20^{\circ}$),
and by the fact that AMS spends 100\% of its time pointing to the sky
(being attached to a gravity-gradient stabilized Space Station), while EGRET typically points
one-third of the time to Earth. 
AMS lacks the low-energy end of EGRET's range due to the curvature
of the electron-positron pair in AMS's magnetic field, which limits the \gray\ 
aperture to $\geq 300$ MeV.
On the other hand, EGRET lacks the high-energy end of AMS's range due to the effect of
electromagnetic backsplash in EGRET's NaI calorimeter which vetos most events above 30 GeV.

The above table shows that \amsg\ and EGRET have very similar \gray\ detection
capabilities, so that AMS will be able to continue and extend the investigation of
galactic and extragalactic \gray\ sources initiated by EGRET.  One main difference
between the two detectors, that of significant aperture above 30 GeV for \amsg,
may lead to the observation of new phenomena in this relatively uncharted region.
A more detailed discussion
of the astrophysics goals of \amsg, and in particular the prospect for novel
discoveries, is given in the next section.  

\section{Gamma Ray Astrophysics Goals of AMS/$\gamma$}\label{astrophysics.sec}

\subsection{Blazar Astrophysics}\label{blazars.sec}

Until the launch of CGRO, only one extragalactic source of $>0.1$ GeV \grays\ 
was known, the blazar 3C 273, detected by the early \gray\ satellite COS-B
\cite{swa78}.
EGRET has subsequently increased the number of detected blazars to over 50
\cite{muk97}, and in fact all of the
extragalactic sources seen by EGRET are blazars.  The term {\it blazar} is used
to describe a wide variety of objects, such as flat-spectrum radio quasars, 
optically violent variable (OVV) quasars, and BL Lac objects, all of which are
characterized by high luminosity and rapid time variability.  Within the ``unified
model'' \cite{ghi98}
these \gray-loud sources have a common source of energy: the
accretion of matter onto a supermassive ($10^{7-10} M_{\odot}$)
black hole, which in turn funnels much of its expelled energy
along the black hole spin axis, forming relativistic ``jets'' of outflowing plasma
\cite{hen91}.
When the axis of such 
a jet points towards Earth, relativistic beaming effects dramatically increase
the apparent luminosity and time variability of these sources, producing a
``blazar'' signature \cite{bla79}

Nearly all EGRET blazars are found to have their luminosity peaking in the
\gray\ region, i.e., at levels higher than the radio, IR, optical, or x-ray regions
\cite{mon95}.
Blazars detected by EGRET are found to belong to the blazar subclasses of
flat-spectrum-radio-quasars and radio-selected BL Lacs \cite{mon95,sam96,ste96b},
whose \gray\ emissions extend to the tens
of GeV, but apparently fall off at higher energies, since they are not
detected by ground-based instruments whose energy thresholds are as low as 200
GeV.  Ground-based instruments, on the other hand, have detected a total of 5
blazars worldwide, all of one subclass,
x-ray-selected BL Lacs.  Of these 5 blazars, EGRET has detected only
one (Mrk 421).  Clearly an interesting transition is occurring between the
effective EGRET cutoff of $\sim 20$ GeV and the present ground-based 
detector threshold
of $\sim 200$ GeV, the observation of which may help determine the \gray\ 
production mechanisms within the blazar jets.  It is not known whether the
spectral cutoffs at $\sim 10^{2}$ GeV of EGRET blazars are due to intrinsic processes
within the jets, or are due to extinction of the higher energy \grays\ off the
diffuse, extragalactic soft photon (IR-optical) background \cite{ste92}
(via $\gamma \gamma \rightarrow e^{+}e^{-}$ interactions) as they propagate over
cosmic distances to Earth.

\subsubsection{Source Counts}\label{counts.sec}

The number of blazars that we expect to detect with \amsg\  can be
estimated from the EGRET blazar population.  During its first six years
of operation (1991 to 1997) EGRET has claimed the detection of 51 blazars
\cite{muk97}.  Different identification methods,
however, have resulted in an EGRET blazar count ranging from 42
\cite{mat97} to roughly 60 \cite{har97}. Of
the 71 unidentified \gray\ sources detected by EGRET, 14 are at galactic
latitudes $|b|>30^{\circ}$ \cite{tho95},
making them likely blazar candidates.

\begin{figure}[h]
\begin{center}
\mbox{\epsfig{file=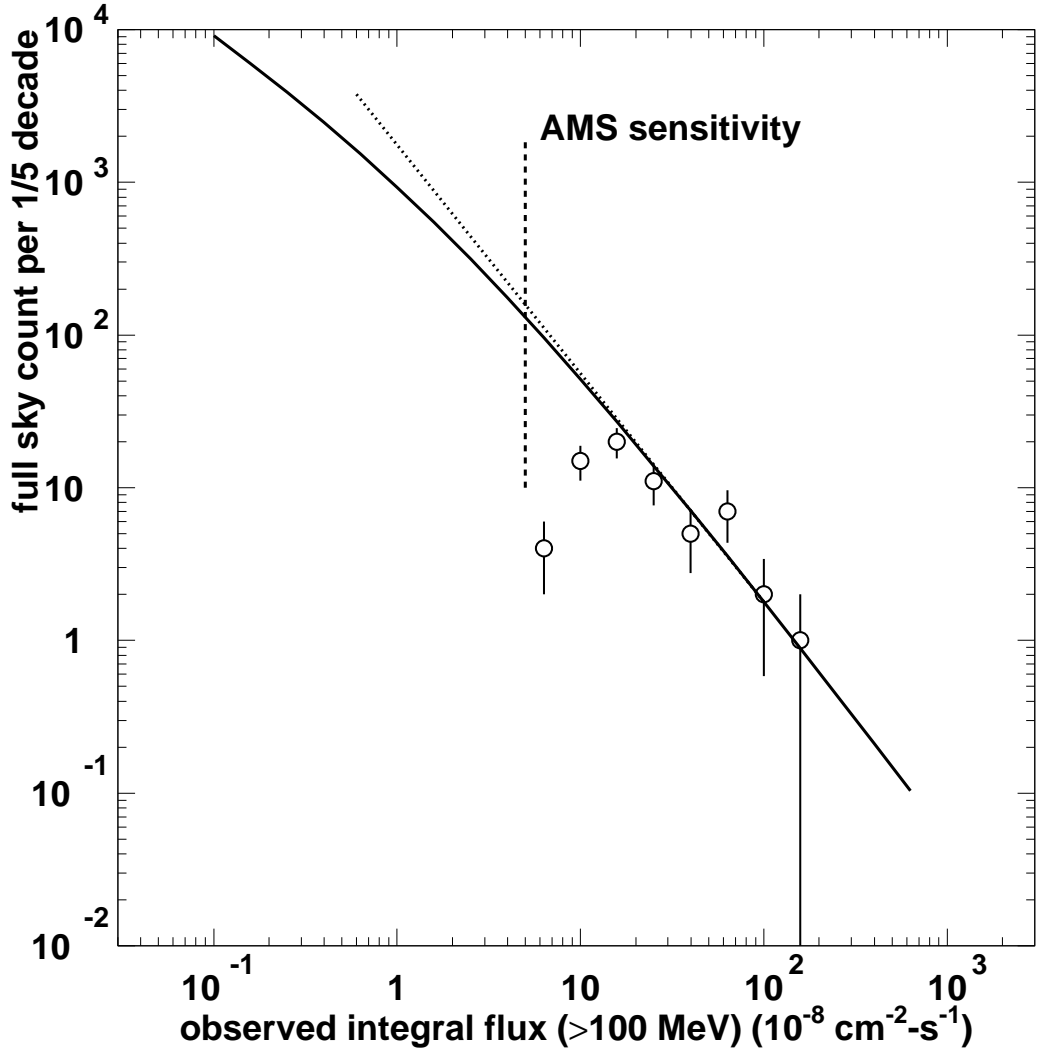,width=6.0in}}
\caption{\small Full sky blazar count as a function of the detected integral
\gray\ flux above 0.1 GeV.  The ordinate gives the total source counts within
{\it one-fifth}-decade intervals of integral flux.  
The unfilled circles represent EGRET detections, taken from their Second
Catalog \cite{tho95}; the solid line is the predicted sky count
from Stecker and Salamon (1996), and the dotted line is the
Euclidean relation, $N(>S)\propto S^{-3/2}$.  The dashed line shows
the point source integral flux sensitivity of \amsg.}
\label{fluxdist-nim.fig}
\end{center}
\end{figure}

%%\begin{wrapfigure}{r}{8.5cm}
%\begin{figure}
%\begin{center}
%\mbox{\epsfig{file=fluxdist99-nim.eps,width=7.7cm}}
%\caption{\small Full sky blazar count as a function of the detected integral
%\gray\ flux above 0.1 GeV.  The ordinate gives the total source counts within
%{\it one-fifth}-decade intervals of integral flux.  
%The unfilled circles represent EGRET detections, taken from their Second
%Catalog \cite{tho95}; the solid line is the predicted sky count
%from Stecker and Salamon (1996), and the dotted line is the
%Euclidean relation, $N(>S)\propto S^{-3/2}$.  The dashed line shows
%the point source integral flux sensitivity of \amsg.}
%\label{fluxdist-nim.fig}
%\end{center}
%%\end{wrapfigure}
%\end{figure}
Figure \ref{fluxdist-nim.fig} shows the flux distribution of these
sources (the identified blazars, plus the 14 high-latitude unidentified
sources).  In this figure the flux is the integral flux above 0.1 GeV,
which is the effective energy threshold for the EGRET blazar studies,
and the effective cutoff seen at $\sim 10^{-7}$ cm$^{-2}$s$^{-1}$ corresponds
to EGRET's quoted flux sensitivity.  The solid line shows the expected
flux distribution of blazar \gray\ sources under the assumption that
the radio and \gray\ fluxes are linearly correlated \cite{ste96,pad93}.
The dashed vertical line
is placed at the estimated flux sensitivity of \amsg, which is about
a factor of two lower than that of EGRET.  Note that we have previously
defined our flux sensitivity in terms of the integral flux above 1 GeV
(not 0.1 GeV), since our detector's energy threshold is effectively at
0.3 GeV. To place the ``equivalent'' \amsg\ sensitivity on this figure
with the EGRET data, we have assumed a nominal $E^{-2}$ differential
source energy spectrum.

From the figure, it is seen
that \amsg\  during its three-year mission should detect somewhere between
120 to 170 blazars, thus increasing the population statistics
of these sources.  This result is relatively model-independent; if
we assume a spatially homogeneous population of sources in Euclidean
space with no evolution, the count $N$ versus flux $S$ distribution 
follows a $N(>S)\propto S^{-3/2}$ power law, as given by the dotted line in
Figure \ref{fluxdist-nim.fig}, independent of the population's luminosity
distribution function.  This curve differs only slightly from the
evolution-dependent curve at the \amsg\  sensitivity.

With these increased statistics, improved \gray\ luminosity distribution
functions can be determined \cite[e.g.,]{chi98},
and compared to those at other wavelengths, particularly at
the radio \cite[e.g.,]{dun90}.
Examination of the correlation between the radio and
\gray\ luminosity functions would help, for example, to discriminate
between \gray\ production models which predict different beam
profiles of the emitted radiation (the \gray\ intensity as a 
function of viewing angle from the jet axis).  For example,
if the photons upscattered
by the synchrotron-emitting electrons are ``external'' photons, coming
from an accretion disk, the beam profile will be substantially narrower
than if the upscattered photons are the synchrotron photons
\cite{der95}.

\subsubsection{Detection of the Extragalactic Soft Photon Background}\label{ib.sec}

All of the blazars detected by EGRET with high ($>6\sigma$) significance
are found to have power law spectra, $dN/dE\propto E^{-\alpha}$, where
the spectral indices $\alpha$ range from 1.7 to 2.5, with a mean index
of 2.15 \cite{muk97}.  The measured energies range from the EGRET
threshold of 0.1 GeV up to several GeV, the value of the upper energy being
statistics limited and hence a function of source strength.  Under the
assumption that one can extrapolate these power law spectra to TeV energies,
several EGRET blazars should be detectable with ground-based atmospheric
Cerenkov telescopes.  In observations of EGRET blazars by the Whipple
Observatory \cite{ker95}, however, {\it no}
EGRET blazars other than the BL Lac object Mrk 421 were detected.  For the
blazars 3C 279, 1633+382, and 2022-077, for example, the extrapolated EGRET
flux is well above the sensitivity limit of the Whipple, yet these sources
are not seen.  Clearly something is happening to the \gray\ spectra of
these blazars between a few GeV and the Whipple threshold energy (400 GeV
at the time of these measurements).

\begin{figure}[h]
\begin{center}
\mbox{\epsfig{file=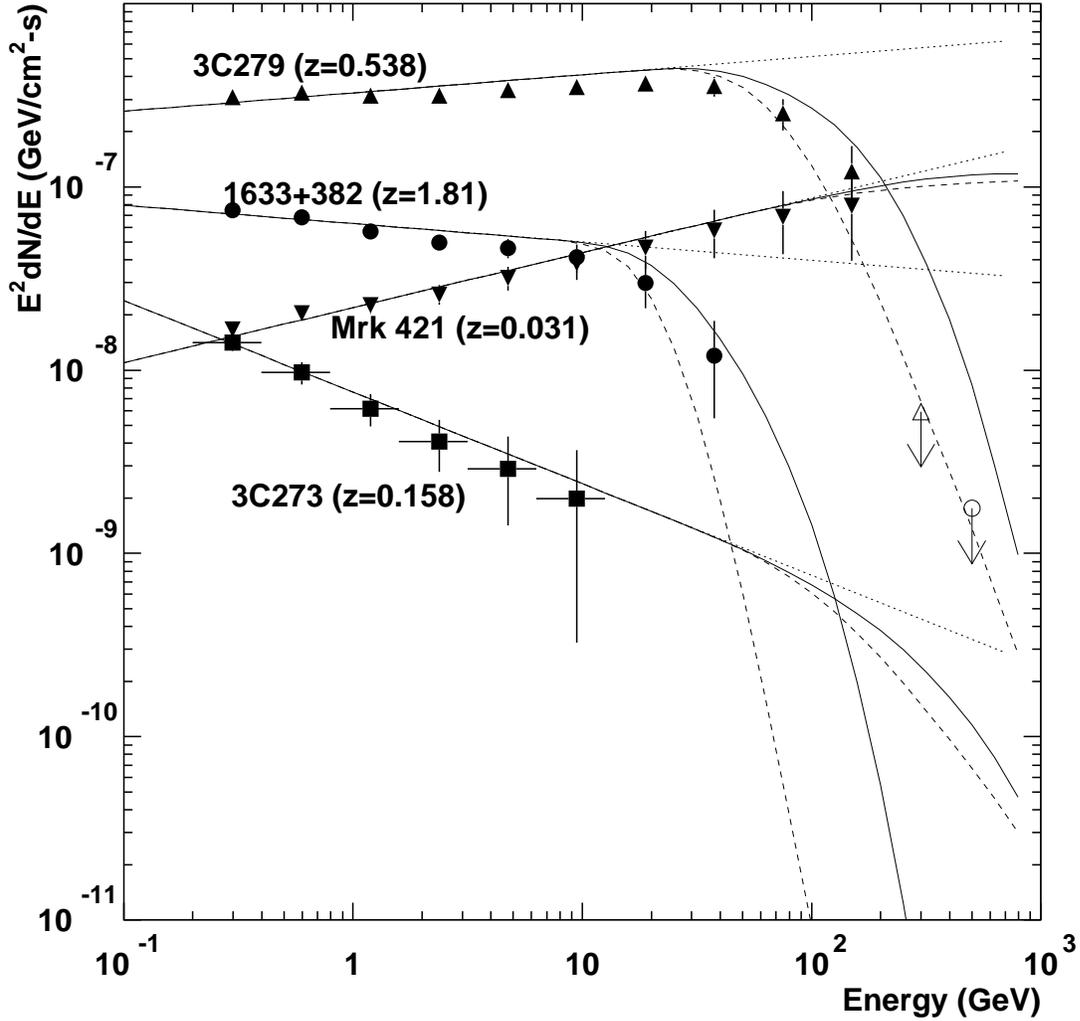,width=6.0in}}
\caption{\small Power law spectra of the blazars 1633+382, 3C273, 3C279, and 
Mrk 421, extrapolated from the measurements of EGRET.  
The redshifts $z$ of these sources are indicated in the figure.
The dotted lines
show the pure power law extrapolation, while the solid and dashed lines
show the effect of \gray\ extinction via pair production off the
intergalactic soft photon (IR to UV) background according to two different
models of the soft photon background \cite{sal98}.
The open triangle and circle represent \gray\ flux upper limits from
3C279 and 1633+382, respectively, as measured by the Whipple Observatory
\cite{ker95}.  These are seen to be well below the naive 
extrapolation of the EGRET power law spectra of these two sources.
Note that we show only {\it one} power-law representation
for each source; some sources, such as 3C279, are highly variable, with
significant shifts occurring in both 
the integrated \gray\ flux and spectral index 
during flare states \cite{muk97}.  Also shown are the
expected \amsg\ spectrum measurements corresponding to a two-year
integration:  3C279 (filled upper triangles), 1633+382 (filled circles),
Mrk 421 (filled lower triangles), and 3C273 (filled squares).}
\label{blazaratt-nim.fig}
\end{center}
\end{figure}

%%\begin{wrapfigure}{r}{8.5cm}
%\begin{figure}
%\begin{center}
%\mbox{\epsfig{file=blazaratt99-nim.eps,width=7.7cm}}
%\caption{\small Power law spectra of the blazars 1633+382, 3C273, 3C279, and 
%Mrk 421, extrapolated from the measurements of EGRET.  
%The redshifts $z$ of these sources are indicated in the figure.
%The dotted lines
%show the pure power law extrapolation, while the solid and dashed lines
%show the effect of \gray\ extinction via pair production off the
%intergalactic soft photon (IR to UV) background according to two different
%models of the soft photon background \cite{sal98}.
%The open triangle and circle represent \gray\ flux upper limits from
%3C279 and 1633+382, respectively, as measured by the Whipple Observatory
%\cite{ker95}.  These are seen to be well below the naive 
%extrapolation of the EGRET power law spectra of these two sources.
%Note that we show only {\it one} power-law representation
%for each source; some sources, such as 3C279, are highly variable, with
%significant shifts occurring in both 
%the integrated \gray\ flux and spectral index 
%during flare states \cite{muk97}.  Also shown are the
%expected \amsg\ spectrum measurements corresponding to a two-year
%integration:  3C279 (filled upper triangles), 1633+382 (filled circles),
%Mrk 421 (filled lower triangles), and 3C273 (filled squares).}
%\label{blazaratt-nim.fig}
%\end{center}
%%\end{wrapfigure}
%\end{figure}
Figure \ref{blazaratt-nim.fig} shows the EGRET power law spectra for
four blazars extrapolated to 700 GeV.  It should be noted that the
majority of EGRET blazars are observed to be variable sources, with
time-varying \gray\ flux amplitudes and spectral indices \cite{muk97}.
We have somewhat arbitrary chosen a single flux amplitude and spectral
index for each source from the Second Catalog \cite{tho95} for
purposes of illustration in Figure \ref{blazaratt-nim.fig}.  Also shown
are \gray\ flux upper limits for 1633+382 and 3C279 measured by the
Whipple ACT, which are seen to be far below the extrapolated power law
spectra for these sources.  

There are a few possible reasons for a spectral cutoff between a few
GeV and a few hundred GeV: (1) a cutoff in the electron spectrum due
to synchrotron and inverse-Compton cooling may cause a corresponding cutoff
in the upscattered \gray\ spectrum (an ``intrinsic'' cutoff mechanism)
\cite[e.g.,]{sik94};
(2) \gray\ extinction through pair-producing interactions with the
diffuse infrared (IR) to ultraviolet (UV) photons of the intergalactic
photon background (IB) (an ``extrinsic'' cutoff mechanism) \cite{gou66,ste92};
and/or
(3) \gray\ extinction through pair-producing
interactions with photons in the local environment of the jet, produced
either within the jet or from the neighboring accretion disk (an ``intrinsic''
cutoff mechanism).

By directly measuring
the spectral cutoffs of the stronger blazar souces such as
1633+382 and 3C279, \amsg\  may determine whether these known cutoffs
are due to intrinsic mechanisms, or to the extrinsic process of interaction
with IB photons.  The main criterion for identifying the ``extrinsic''
mechanism is that the cutoff energies should decrease with increasing
source redshift $z$, since the integrated opacity of the IB increases with
source distance. 
The opacity $\tau$ to a source of redshift $z_{e}$
at a measured \gray\ energy of $E_{0}$ is given by
\begin{equation}
\tau(E_{0},z_{e})=c\int_{0}^{z_{e}}dz\,\frac{dt}{dz}\int_{0}^{2}dx\,
\frac{x}{2}\int_{0}^{\infty}d\nu\,(1+z)^{3}\frac{u_{\nu}(z)}{h\nu}
\sigma_{\gamma\gamma}(s),
\label{tau.eq}
\end{equation}
where $u_{\nu}$ is the differential IB energy density at frequency $\nu$,
$\sigma_{\gamma\gamma}(s)$ is the $\gamma\gamma\rightarrow e^{+}e^{-}$
cross section as a function of the center-of-mass energy $\sqrt{s}$,
$t$ is time,
and $x\equiv(1+\cos\theta)$, $\theta$ being the angle between the \gray\
and the soft photon momenta.
Figure \ref{blazaratt-nim.fig} shows the calculated \gray\ 
attenuation for two different models of the near-IR to UV intergalactic
photon background \cite{sal97};
one sees that the cutoff energy decreases with increasing redshift of the
source. This figure also shows the ``fauxdata'' points of the \amsg\ 
spectra for these sources acquired over a two-year period of operation,
under the assumption (for illustrative
purposes) of power law spectra that are constant in time.  These are obtained
by convolving the source spectrum with the instrumental energy resolution
function shown partially in Figure \ref{energyres.fig}.
The energy bin
widths are indicated in the fauxdata for 3C273; the statistical errors
are also shown at each energy point for the four sources.
The effects of attentuation (according to the solid line predictions) 
have been included in the calculation of the fauxdata.

Cutoffs are evident in the \amsg\ spectra of the stronger
sources at higher redshifts (3C279 and 1633+382), at energies
ranging from 30 to 100 GeV.  At higher energies the statistics are too low
(and the \amsg\  energy resolution is significantly worsened) for
AMS to measure blazar cutoffs.   Since the threshold condition for
pair production is $2E\epsilon(1-\cos\theta)>4m_{e}^{2}c^{4}$, where
$\epsilon$ is the background photon energy, for a \gray\ of energy
$E_{\rm TeV}$ only those background photons with energy 
$\epsilon_{\rm eV}>0.3/E_{\rm TeV}$ can contribute to the opacity.
Thus \amsg\ will only be able to detect \gray\ attentuation due
to the IB of optical and UV photons.  Therefore the observation of
such a redshift-dependent cutoff would constitute a {\it direct}, finite
measurement of
the intergalactic optical and UV photon background, which has not yet
been made.  Conversely,
the {\it absence} of a redshift dependence for observed blazar cutoffs
would definitively point to an intrinsic cutoff mechanism in the \gray\ 
production in jets.

Finally, it has also been pointed out that if the IB spectrum 
$u_{\nu}(z)$ were known to high precision, then a measurement of
blazar cutoffs, should they be extrinsic, would provide a measure of
Hubble's constant $H_{0}$ \cite{sal94},
since the $dt/dz$ term in Eq. \ref{tau.eq}
contains a factor of $H_{0}^{-1}$.  Unfortunately, it is unlikely that
such accurate spectral information will be available in the forseeable future
in the optical or UV bands.

\subsection{Gamma Ray Bursts}\label{grb.sec}

Gamma ray bursts (GRBs) are likely to be the most energetic phenomena that occur
in the universe.  These point-source events flare into being with short, intense bursts 
of x-rays and \grays, then quickly fade away.  GRBs are isotropically distributed in
the sky, with no statistical evidence for repeating bursts occuring from a single source.
First detected in the 1960's by the Vela satellites (whose function it was to
search for nuclear weapons testing), it is only within the last
two years that this class of events has been identified as being cosmological in origin,
implying energy outputs on the order of $10^{51}$ to $10^{53}$ ergs released within
a time scale of seconds.  This rivals the mean luminosity of {\it the entire universe},
approximately $10^{53}$ erg/s \cite{hur98}.  

One of the two most popular current models for GRBs is the merger of binary neutron stars
(with a maximum energy release of $5\times 10^{53}$ ergs) or of a neutron star and
black hole companion, an event which is estimated to occur on the order of once every
$10^{6}$ years per galaxy, in excellent agreement with the observed rate of GRBs
(about one per day by the BATSE instrument of CGRO) \cite{pir98}.  
The contending model is that GRBs are the results of ``hypernovae'',
extremely energetic supernovae that result from the gravitational collapse of
very massive stars.

The nonthermal x-ray and \gray\ radiation observed from GRBs is most likely produced
by ultra-relativistic particles (electrons and protons) that are accelerated in the
shock waves generated in the expanding fireball that follows the progenitor event.
The strongest detections of GRBs are in the x-ray region of hundreds of keV, in which
the BATSE, BeppoSAX, ROSAT, and ASCA orbiting instruments operate.  Some of the very
strongest bursts have also been detected in GeV \grays\ by EGRET (four bursts
with \grays\ above 0.1 GeV), one GRB yielding
photon energies up to 26 GeV \cite{hur94}.  Given the relatively low sensitivity of
EGRET to GRBs, the detection rate by EGRET is consistent with {\it all} GRBs having
spectra that extend to GeV energies (and possibly higher), possibly
with nearly constant values of $E^{2}dN/dE$ \cite{din97}.

It is in fact the {\it highest} energy \grays\ that are the most constraining on the
fireball models of GRBs; with the significantly larger sensitivity to $>10$ GeV
\grays\ that \amsg\ has compared to EGRET, the timing and spectral data of GRB
\grays\ detected by \amsg\ could be invaluable in elucidating the nature of the
GRB fireball.  Such data could also be applied to the question of whether the
ultra-high energy cosmic rays might originate in the fireball shocks of GRBs \cite{wax95}.
The fundamental issue involves the ``compactness'' of the GRB fireball, which is a
dimensionless ratio of source luminosity to source radius.  The higher the compactness,
the greater the opacity for \gray\ emission, due to the large density of
lower energy photons in the fireball that are above threshold for pair 
production.  Fireball dimensions are causally limited by x-ray timing variations, whose
time constants are sometimes as low as $\sim 10^{-3}$ s.  Based on distance estimates from
optical measurements, the luminosity of the fireballs can be inferred.  These combine
to form an average optical depth for $\gamma\gamma\rightarrow e^{+}e^{-}$ of
$\sim 10^{15}(E/10^{51}{\rm ergs})(\delta T/10 {\rm \ msec})^{-2}$ \cite{pir98}.
In spite of the enormous estimated opacities, \grays\ are observed from GRBs.
This conflict can be resolved if the fireball is expanding with a very large
Lorentz factor.  Relativistic kinematics then both boost the 
\gray\ energy and observed luminosity
of the source, plus reduce the time constants of timing variations, so that the apparent
optical depth is approximately a factor $\Gamma^{4+2\alpha}$ larger than the
actual value ($\alpha$ is the spectral index of the GRB spectrum).  This argues
for bulk Lorentz factors $>10^{2}$.  The higher the observed \gray\ energies, the
larger the fireball Lorentz factor must be to allow these \grays\ to escape the
fireball region \cite{bar97a}.

In addition, there is a relationship between the redshift of a given GRB and
the maximum energy \gray\ that can be received at Earth, due to the effect
of extinction via pair production off the cosmic optical and UV photon background
(see Sec.\ref{ib.sec}).  No photons with energies above 20 GeV, for example, have
been detected from sources {\it known} to have redshifts above unity.  Detection
by \amsg\ of any \gray\ energies above $\sim20$ GeV, coupled with an independent
lower limit to the redshift of the GRB source via optical measurements, would
place limits on the energy densities of the cosmic photon backgrounds if the
redshift were on the order of unity or greater.

The time distribution of the \gray\ events, relative to the prompt x-ray bursts
that are observed, is also of great importance.  In one GRB, GeV \grays\ were
observed by EGRET {\it for over one hour} after the initial x-ray burst \cite{hur94}.
A number of scenarios have been proposed to explain this delayed emission, ranging
from generation in more distant (external) shocks by the fireball, to the deflection
of $e^{+}e^{-}$ pairs in the intergalactic magnetic field, followed by inverse
Compton scattering that produces the \grays\ \cite{pla95}.

It is not possible to predict with certainty how many GRBs will be detected by
\amsg.  Given the comparable total exposures ($>10^{11}$ cm$^{2}$-s-sr for AMS) 
of EGRET and \amsg\ over the lifetimes of the two instruments, one could use the
limited statistics of EGRET's four GeV GRBs as a naive estimate.  In fact, it is
likely that fewer would be detected by \amsg\ due to the higher effective energy 
threshold of 0.3 GeV for \amsg, compared to 30 MeV for EGRET.  A GRB similar to the 
one strong
burst observed by EGRET, GRB940217 \cite{hur94}, would certainly be
detected by \amsg\ if the burst were within a few tens of degrees from the
orbital plane of the Space Station. Since the FOV of EGRET is smaller than that of
\amsg, and since EGRET was not targeting GRB940217 when it occurred, it is 
safe to say that the {\it a priori} detection probability of a strong GeV burst
by \amsg\ would be larger than for EGRET.

\subsection{The Extragalactic Gamma Ray Background at High Energies}\label{egrb.sec}

\begin{figure}[h]
\begin{center}
\mbox{\epsfig{file=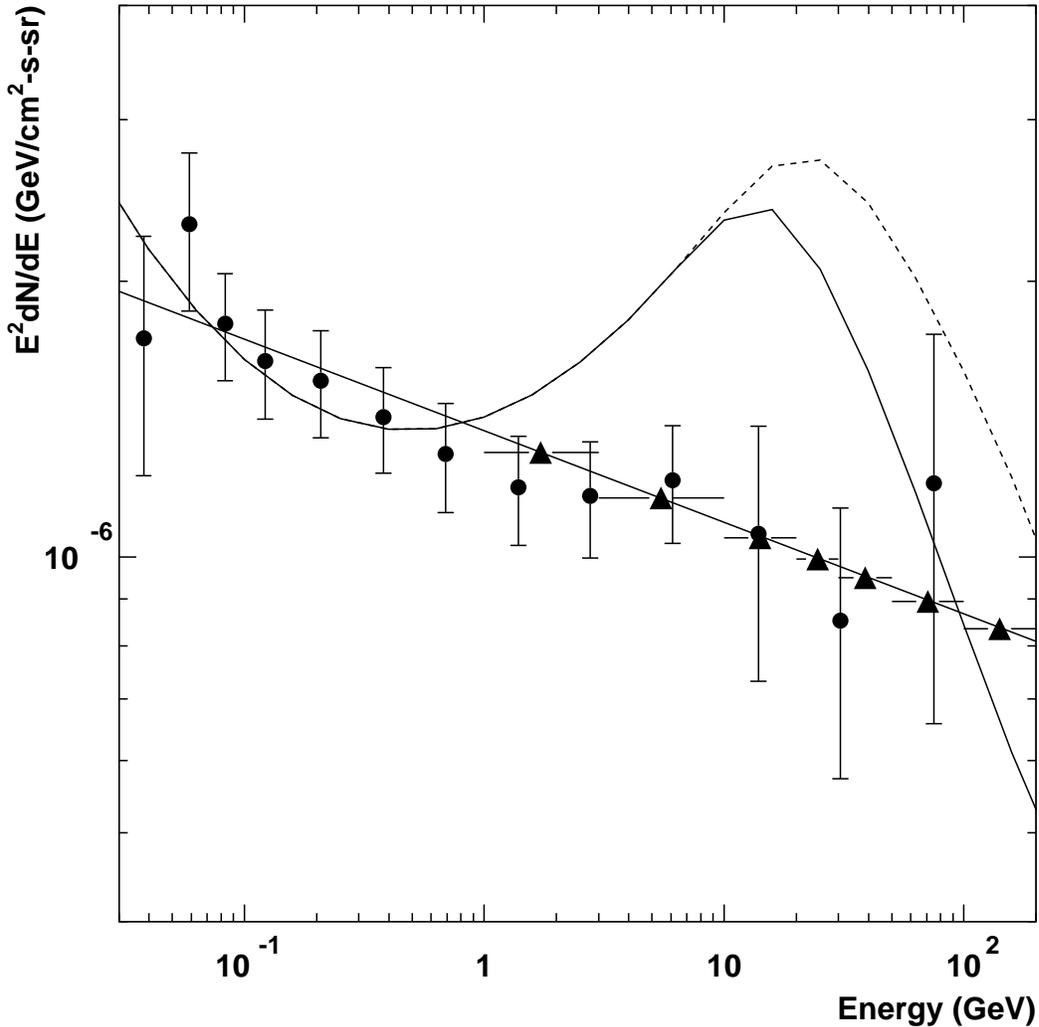,width=6.0in}}
\renewcommand{\baselinestretch}{1.0}
\small\normalsize
\caption{\small The extragalactic gamma ray background as measured by EGRET
(filled circles), along with their fitted (straight solid line) power law
spectrum.  Also shown are two versions of 
a theoretical spectrum calculated under the assumption that 
100\% of the EGRB is due to unresolved blazars \cite{ste96,sal98},
which includes the effects of \gray\ attenuation via interactions with the
IB optical-to-UV photons.  (The solid and dashed line curves differ in
assumptions regarding the effects of evolving cosmic metallicity on the
optical and UV photon backgrounds.)  The peaks in $E^{2}dN/dE$ at $\sim 20$ GeV
in these theoretical spectra 
are sensitive to assumptions regarding the spectral index distribution
of blazars, which appear to be inconsistent with the EGRET data, but could be
readily modified to be more so.  The one feature which is {\it not} malleable
is the spectral cutoff at $\sim 20-30$ GeV; for an EGRB of blazar-dominated origin 
a spectral cutoff in the 20-50 GeV region must exist.  Should the EGRB spectrum
remain a power law out to 100 GeV, the blazar-origin hypothesis would be ruled out.
The error bars on the EGRET data (filled circles) are mostly
systematic (13\% for 0.03 to 10 GeV data; 30\% above 10 GeV)
[Sreekumar etal 1998], representing over 3 years of EGRET data.  AMS 
{\it statistical} errors, based on one year of operation,
are smaller than the data point symbols (filled triangles).  For example,
between 50 to 100 GeV, AMS should see $1.3\times 10^{3}$ EGRB \grays\ in one
year, assuming the EGRET EGRB spectrum. }
\label{egrb-nim.fig}
\end{center}
\end{figure}

EGRET has recently measured the spectrum of the isotropic 
extragalactic \gray\ background
(EGRB) \cite{sre98}, as
shown in Figure \ref{egrb-nim.fig}.  Originally detected by the
SAS-2 satellite telescope at energies above 100 MeV \cite{tho82},
the origin of this background is still
uncertain, and may be due 
to diffuse processes such as secondary production from cosmic ray
collisions with intergalactic gas \cite{dar95,ste96a}, or, more
likely, the accumulated emission of unresolved point sources, such as
blazars or other classes of active galactic nuclei \cite{big79,kaz83}.

\subsubsection{The Role of Blazars}\label{egrb.blazars.sec}

In particular, the hypothesis that unresolved \gray\ blazars are responsible
for the EGRB has been examined, with conflicting conclusions \cite{ste96,chi98}.
Stecker and Salamon (1996), by assuming a linear relation between the radio
and \gray\ luminosity functions of blazars \cite{pad93}, fit EGRET's
blazar flux distribution (see Section \ref{counts.sec}), and also predicted
the unresolved blazar contribution to the EGRB, shown in Figure \ref{egrb-nim.fig}.
They were able to account for 100\% of the observed EGRB, but predicted a sharp
cutoff near 20-40 GeV due to the effects of \gray\ extinction off the IB
optical and UV photons \cite{sal98}.  A more recent estimate of the \gray\ luminosity
function of blazars, however, finds that unresolved blazars contribute
only 25\% of the total EGRB, thus indicating the need for additional (nonblazar)
sources. Such a conclusion would not be inconsistent with an EGRB whose spectrum
continues as a power law to energies above 100 GeV.

Whether blazars can or cannot contribute 100\% of the EGRB
may be resolved with the increased statistics at the highest
energies that \amsg\  will obtain.  If
EGRET-type blazars
do contribute 100\% of the EGRB, then a cutoff {\it must} be present somewhere
below 100 GeV.   Although
the latest EGRET data shows no evidence of a cutoff in the EGRB spectrum,
and is consistent with a pure power law (Eq.\ref{egrb.eq}) \cite{sre98},
EGRET's systematic errors make the current data somewhat inconclusive.
\amsg\  in two years of operation should detect $\sim10^{5}$
EGRB \grays\ in the interval 0.5-1.0 GeV, and $\sim10^{3}$ between
50-100 GeV; assuming that systematics can be kept to the several percent
level, a spectral cutoff occurring within the 20-40 GeV region would be detectable
by \amsg.

\subsubsection{Instrumental Backgrounds to the EGRB Measurement}
\label{egrb.backg.sec}

Any measurement of the EGRB requires that instrumental background events
be kept to a lower rate than the EGRB flux.  To estimate the effects
of background due to collisions of cosmic ray electrons, positrons, and
protons with the AMS instrument, we divided the \gray\ spectrum into
four bins per decade of energy (from 0.5 to 100 GeV) and required that
we investigate backgrounds down to a level of 20\% of the EGRB rate
in {\it each} energy bin.  For example, in the 25-40 GeV bin it will
take \amsg\ $2.0\times 10^{5}$ seconds to obtain 5 EGRB \grays.  In our
Monte Carlo analysis, therefore, we threw $2.0\times 10^{5}$ seconds' worth of
cosmic ray electron, positron, and proton flux at the instrument to
generate a false \gray\ background.  These cosmic rays were thrown isotropically
over a zenith angle range of $0^{\circ}<\theta <110^{\circ}$, where the
largest zenith angle corresponds to the location of the Earth limb at
an orbital altitude of 400 km.  The energies were distributed according
to known electron, positron, and proton energy spectra as discussed below.

For cosmic ray electrons above 10 GeV, we used the compiled data presented
in M\"{u}ller and Tang (1987) \cite{mul87}
to estimate the electron spectrum, and for energies below
10 GeV we used the data of Golden et al. (1994) \cite{gol94}.
We note that this latter set of data was obtained by
a balloon flight in northeastern Canada, where the geomagnetic rigidity
cutoff is well below 1 GV; since AMS during most of its orbit will be in
regions with much higher rigidity cutoff (up to 15 GV near the equator),
the use of the Golden etal data for the low energy electron spectrum is
quite conservative.  Approximately $10^{7}$ electrons in all were thrown
at the AMS instrument, of which 0.1\% met the basic reconstruction criteria
for a \gray\ candidate event.  Additional quality cuts were applied so that
this background was reduced to at most one count per energy bin
(for a total of 2 background events).  These
cuts were then applied to the actual \gray\ data as described in
Section \ref{mc.sec}.  The positron flux was assumed to be 0.1 times
that of the electron flux over the full spectrum (measurements give
this ratio as varying between 0.05 to $\sim 0.2$ above 0.5 GeV \cite{bar97}).
For a given \gray\ energy bin
[$E_{\rm low},E_{\rm high}$], only electrons and positrons of energies
greater than $E_{\rm low}$ were considered to be potential sources of 
\gray\ background.  This assumption, justified by the fact that none
of the reconstructed background events had energies higher than their
progenitor electrons or positrons, reduced computational demands 
significantly.

For cosmic ray protons we used the recent absolute flux measurements
of Menn etal (1997) \cite{men97}
which were obtained in a balloon flight at low geomagnetic cutoff.  As with
the electrons, therefore, our estimated low energy proton flux is
conservatively high.  Unlike electrons, which can generate bremsstrahlung
photons with nearly the full primary electron energy, protons collisions
generate \grays\ primarily through $\pi^{0}$ decays (direct
\gray\ production is only at the level of 10\% that of $\pi^{0}$s
at 500 GeV/c, for example \cite{alv93}),
where the secondary
$\pi^{0}$s carry only a fraction of the primary proton energy.  Therefore,
even though the CR $p/e$ ratio is $\sim 10^{2}$, we do not need to
follow $10^{9}$ protons (compared to $10^{7}$ electrons) through the detector.
Instead, when estimating background production for the \gray\ energy
bin [$E_{\rm low},E_{\rm high}$], only protons of energy $>\eta E_{\rm low}$,
$\eta>1$, were considered.  We chose the value $\eta=7$ based on calculations
of galactic \gray\ production via CR proton interactions with the interstellar
medium by Mori \cite{mor97}.  Using the HADRIN, FRITIOF, and
PYTHIA particle collision codes, and convolving with the CR proton 
spectrum, Mori found that for a fixed \gray\ energy $E_{\gamma}$ 
roughly half the photons were produced by protons of energy $>7E_{\gamma}$.
This reduced our computational load by $\sim\eta^{-1.7}$ given the integral
proton spectrum of $E^{-1.7}$ above a few GeV; approximately $3\times10^{7}$
protons were required for these studies.  We found that with the quality
cuts introduced to eliminate the electron-induced background events, we
had at most one proton-induced background \gray\ per source-energy bin (for a total
of 5 background events); all these had reconstructed energies of $<1$ GeV.

\subsection{Detection of Neutralino Annihilation}\label{neutralino.sec}

\begin{figure}[h]
\begin{center}
\mbox{\epsfig{file=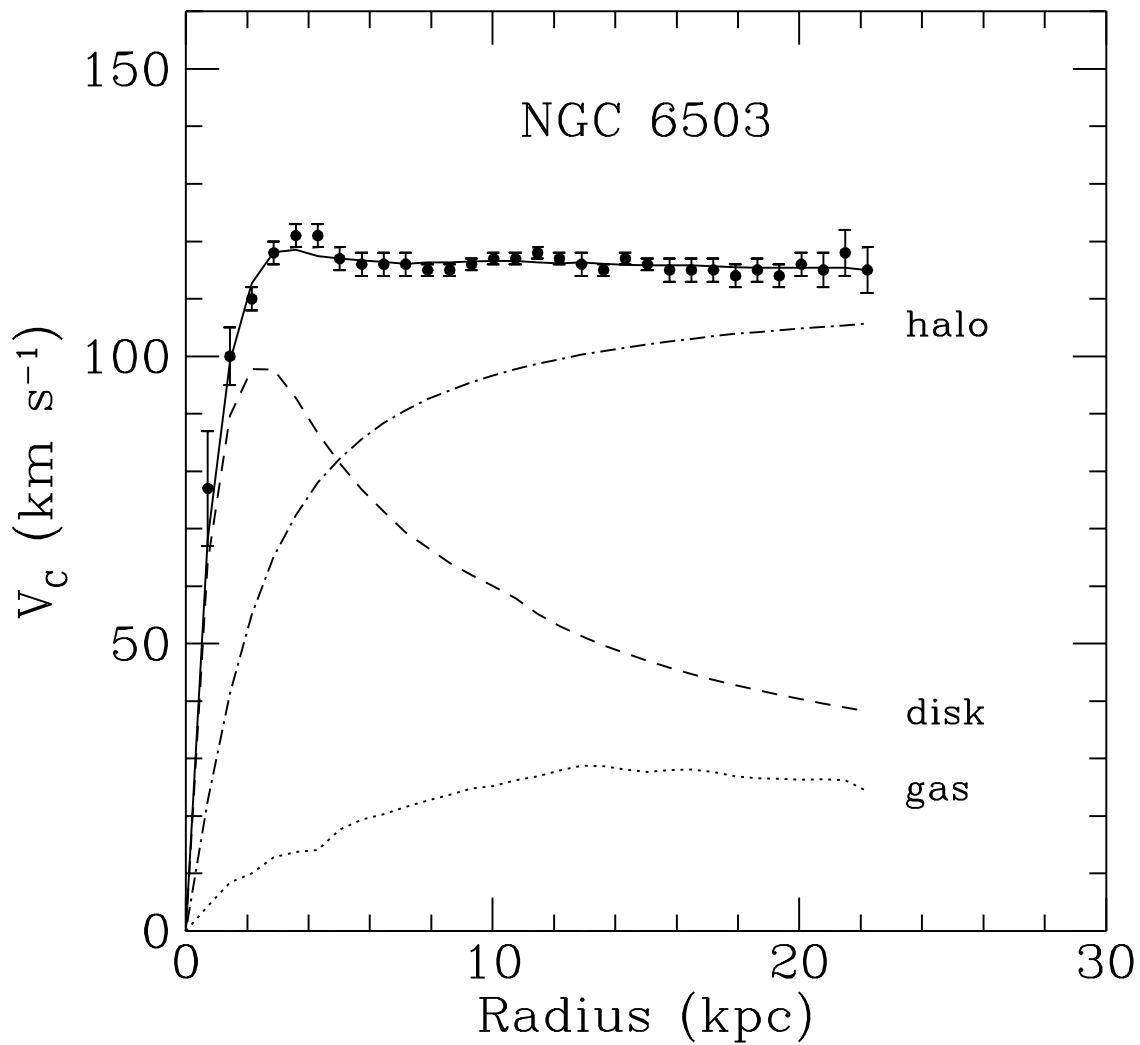,width=6.0in}}
\renewcommand{\baselinestretch}{1.7}
\small\normalsize
\caption{\small The rotation curve of the 
galaxy NGC 6503 is shown as a function of radius from
the galaxy center, as measured by Doppler shifting of the 21-cm radio line from gaseous
HI in the disk.  The flatness of the rotation curve at large radii implies an integrated
galaxy mass $M(r)$ that falls off roughly as $r^{-1}$, inconsistent with the observed
mass profiles of the visible disk and gas components of the galaxy.  The inclusion of
a dark matter halo component with density profile $\sim r^{-2}$ provides an excellent fit to
the observed rotation curve.
This figure is taken from Begeman, Broeils, and Sanders (1991)\cite{beg91}.}
\label{rotation-nim.fig}
\end{center}
\end{figure}

It is well known that the luminous matter within galaxies, and within our Galaxy in
particular, is far below that required to explain the gravitational dynamics of their
components.
%%\begin{wrapfigure}{r}{8.5cm}
%\begin{figure}
%\begin{center}
%\mbox{\epsfig{file=rotation-nim.eps,width=7.7cm}}
%\caption{\small The rotation curve of the 
%galaxy NGC 6503 is shown as a function of radius from
%the galaxy center, as measured by Doppler shifting of the 21-cm radio line from gaseous
%HI in the disk.  The flatness of the rotation curve at large radii implies an integrated
%galaxy mass $M(r)$ that falls off roughly as $r^{-1}$, inconsistent with the observed
%mass profiles of the visible disk and gas components of the galaxy.  The inclusion of
%a dark matter halo component with density profile $\sim r^{-2}$ provides an excellent fit to
%the observed rotation curve.
%This figure is taken from Kamionkowski (1998).}
%\label{rotation-nim.fig}
%\end{center}
%%\end{wrapfigure}
%\end{figure}
The inability of the estimated mass from 
disk and gas components to account for the observed rotation curves of spiral
galaxies, for example, is exemplified in Figure \ref{rotation-nim.fig}, 
implying the presence of
a ``dark matter'' component whose density falls off roughly as $r^{-2}$, where $r$
is the distance from the center of the galaxy.  Although MACHOs 
(massive compact halo objects) may possibly account for a
significant fraction (20\% to 100\%) of the inferred dark matter in our Galaxy
\cite{alc97,sut98}, the role of MACHOs in the mass budget of the Galaxy is still very poorly
defined, and may prove to be minor \cite{fuk98}.  An alternative is that
this dark component is comprised of WIMPs (weakly interacting massive particles), one
candidate for which is the supersymmetric neutralino, $\chi$.  (The subject of
supersymmetric dark matter has recently been comprehensively reviewed by 
Jungman, Kamionkowski, and Griest (1996)\cite{jun96}.)

Should supersymmetric neutralinos be the dominant dark matter component, the
mutual annihilation of these Majorana particles near the center of the galaxy would
produce a number of signatures, one of which would be the emission of both line and
continuum \grays\ whose characteristic spectra would distinguish them from other
contributions to the diffuse \gray\ background.  The flux of monochromatic (line) \grays\ 
per solid angle from $\chi\chi\rightarrow\gamma\gamma$ or $\chi\chi\rightarrow Z\gamma$
annihilations in the galactic halo is given by
\cite{tur86}
\begin{equation}
\label{dfdo.eq}
\frac{dF}{d\Omega}=\eta_{\gamma}\frac{\left\langle\sigma_{\chi\chi}v\right\rangle}
{8\pi M_{\chi}^{2}}\int_{0}^{\infty}\rho_{\chi}^{2}(l,\psi)\,dl(\psi),
\end{equation}
where $\rho_{\chi}$ is the neutralino mass density, $M_{\chi}$ is the neutralino mass,
$\sigma_{\chi\chi}$ is the annihilation cross section into $\gamma\gamma$ or $Z^{0}\gamma$,
the leading factor $\eta_{\gamma}=2$ for $\chi\chi\rightarrow\gamma\gamma$ and
$\eta_{\gamma}=1$ for $\chi\chi\rightarrow Z\gamma$,
$\psi$ is the angle between the viewing direction and the direction to the center of the
Galaxy, and $l$ is the distance along the viewing direction.
Because the annihilating particles are identical, the
production rate per unit volume is 
$\eta_{\gamma}(n_{\chi}^{2}/2)\left\langle\sigma_{\chi\chi}v\right\rangle$,
where the factor of 1/2 in $n_{\chi}^{2}/2$ eliminates double-counting
of annihilating pairs.  Equation \ref{dfdo.eq} can be expressed in the form
\begin{equation}
\label{dfdo2.eq}
\frac{dF}{d\Omega}=\left[3.16\times10^{-11}{\rm cm}^{-2}{\rm s}^{-1}{\rm sr}^{-1}\right]
\left[\frac{\eta_{\gamma}}{2}\right]
\left[\frac{\left\langle\sigma_{\chi\chi\rightarrow\gamma\gamma}v\right\rangle}{10^{-29}
{\rm cm}^{3}{\rm s}^{-1}}\right]\left[\frac{10 {\rm GeV}}{M_{\chi}}\right]^{2}
\left[\frac{\rho_{0}}{0.4 {\rm GeV/cm}^{3}}\right]^{2}
\left[\frac{R_{0}}{8.0 {\rm kpc}}\right]J(\psi),
\end{equation}
where 
\begin{equation}
J(\psi)\equiv \int_{0}^{\infty}\frac{\rho_{\chi}^{2}}{\rho_{0}^{2}}\,
d\left(\frac{l}{R_{0}}\right)(\psi).
\label{jpsi.eq}
\end{equation}
Estimates of the distance of the Sun to the Galactic center, $R_{0}$, vary from
6 to 10 kpc, with a weighted mean of $8.0\pm 0.5$ \cite{rei93} (to be compared with
the ``standard'' value of 8.5 kpc adopted by the IAU in 1985 \cite{fic91}). We take
$R_{0}=8.0$ kpc.
The local dark matter density $\rho_{0}$ is obtained from measurements of the circular rotation 
velocity $v_{c}(R)$ of our Galaxy, where $R$ is the galactocentric radius. The 
contributions to $v_{c}(R)$ from
the disk and bulge (baryonic) components are subtracted, leaving a residual of
$v_{c,halo}(R)^{2}\approx GM(R)/R$.  Fitting this to various dark matter
halo profiles then yields the parameter $\rho_{0}$.  This has been done recently
by Dehnen and Binney \cite{deh96}, who obtain values of $\rho_{0}$ ranging from 
0.1 to 0.7 GeV/cm$^{3}$, depending on the assumed halo density profile.  These authors
also conclude that the mass distribution of the Galaxy is currently ill-determined, so
no one model is clearly preferable over others.

The form of the halo density profile $\rho(R)$ is critical to the question of observability
of \grays\ from neutralino annihilation.  The simplest
assumption, that of an isothermal halo, leads to a density profile $\rho$ which is constant
for $R\ll a$, where $a$ is the ``core'' radius, and which falls as $r^{-2}$ for $R\gg a$
\cite{bin87}, accounting for the flat rotation curves of spirals at large radii.
High resolution, N-body
numerical simulations of the collapse of cold dark matter (CDM) halos, however, show no signs
of the existence of a halo ``core'' down to the smallest resolvable scales.
Studies by Navarro, Frenk, and White (NFW) (1996,1997)\cite{nav96,nav97}, 
corroborated by others \cite{col95}, show
that CDM halos over a wide range of total mass obey a ``universal'' density profile, given by
\begin{equation}
\rho(R)\propto \frac{1}{R(1+R/a)^{2}},
\end{equation}
where $a$ is a scale radius.  Although this profile results in a singular annihilation rate 
at the galactic center, the total \gray\ flux over any finite solid angle about $\psi=0$ is 
finite.  Even steeper radial dependences have been proposed:  Berezinsky, Gurevich, 
and Zybin (1992)\cite{ber92}
argue theoretically for $\rho(R) \propto R^{-1.8}$ down 
to a cutoff of $R\sim 1$ pc, but this result
has been criticized as arising from their use of unphysical initial conditions \cite{flo94}.
However, in the mass modelling of the Galaxy by Dehnen and Binney \cite{deh96}, one of their models
has an $R^{-1.8}$ density profile down to a core radius of $a=1$ kpc, and fits the dynamical
data with a good $\chi^{2}$.
Recent analyses of dwarf and low-surface brightness galaxies, whose dark matter fraction may
reach levels of 95\%, indicate the presence of central cusps with density profiles
$\rho(R)\propto 1/R^{\gamma},\ \gamma\approx 0.2-0.4$ \cite{kra98}, 
significantly shallower than that of NFW.  It is not clear, however, that these shallower
cusps are present in larger galaxies, such as our Galaxy, since the cusp steepness may depend
on how dynamically ``relaxed'' the halo has become \cite{kra98}.

To estimate the potential sensitivity of AMS to monochromatic photons from neutralino
annihilation, we use three of the halo profiles determined by Dehnen and Binney,
corresponding to  their models 2d, 4d, and 2f, plus a generic 
quasi-isothermal profile.  All the Dehnen and Binney halo profiles
take the form
\begin{equation}
\rho(\varrho)=\rho_{a}\left(\frac{\varrho}{a}\right)^{-\gamma_{h}}
\left(1+\frac{\varrho}{a}\right)^{\gamma_{h}-\beta_{h}},
\end{equation}
where $\varrho=(R^{2}+z^{2}/q^{2})^{1/2}$, $R$ being the radial distance along the plane of
the galaxy, $z$ being the height above the plane, and $q=0.8$ being the assumed oblateness 
of the halo.  The halo profile exponents $\gamma_{h}$ and $\beta_{h}$ vary with the mass
model, and are given in table below:
\begin{center}
\begin{tabular}[h]{|c|c|c|c|c|c|}\hline
model & $\gamma_{h}$ & $\beta_{h}$ & $a$ (kpc) & $\rho_{a}$ ($M_{\odot}$/pc$^{3}$) 
	& $\rho_{0}$ (GeV/cm$^{3}$)\\ \hline
D\&B 2d & 1 & 3 & 21.8 & 0.006159 & 0.34 \\
D\&B 4d & 1 & 3 & 5.236 & 0.1101 & 0.43 \\
D\&B 2f & 1.8 & 1.888 & 1.0 & 0.4179 & 0.31 \\
isotherm & & & 1.0 & & 0.30 \\ \hline
\end{tabular}
\end{center}

\begin{figure}[h]
\begin{center}
\mbox{\epsfig{file=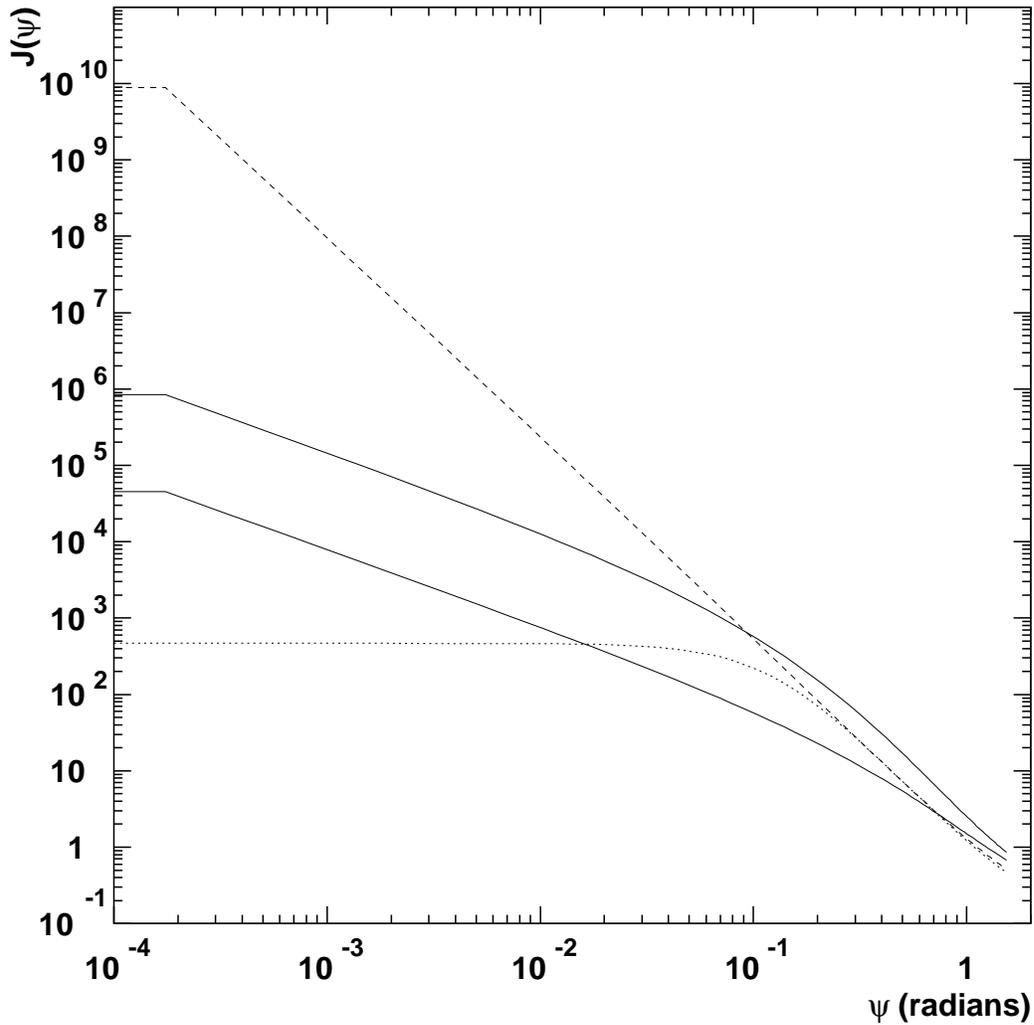,width=6.0in}}
\caption{\small The integrated \gray\ flux $J(\psi)$ versus the angle $\psi$ from the
Galactic center for the tabulated halo dark matter density profiles, where $J(\psi)$ is
defined in Eq. \ref{jpsi.eq}.  The dashed, upper and lower solid, and dotted lines 
correspond respectively to D\&B models 2f, 4d, 2d, and the quasi-isothermal halo profile.
The most optimistic profile is model 2f, which is very close to that of
Berezinsky et al (1992).  Models 4d and 2d are both NFW halo profiles.}
\label{jpsi-nim.fig}
\end{center}
\end{figure}

In models 2d and 4d, the
density profile is fixed to follow the ``universal'' profile found by NFW, for which
$\gamma_{h}=1$ and $\beta_{h}=3$.  The last halo profile is an approximation to the
true isothermal profile (see Binney \& Tremaine, 1987):
\begin{equation}
\rho(R)=\rho_{0}\left(\frac{a^{2}+R_{0}^{2}}{a^{2}+R^{2}}\right).
\end{equation}
Figure \ref{jpsi-nim.fig} shows the resulting $J(\psi)$ curves for the various halo profiles.

%%\begin{wrapfigure}{r}{8.5cm} 
%\begin{figure}
%\begin{center}
%\mbox{\epsfig{file=jpsi99-nim.eps,width=7.7cm}}
%\caption{\small The integrated \gray\ flux $J(\psi)$ versus the angle $\psi$ from the
%Galactic center for the tabulated halo dark matter density profiles, where $J(\psi)$ is
%defined in Eq. \ref{jpsi.eq}.  The dashed, upper and lower solid, and dotted lines 
%correspond respectively to D\&B models 2f, 4d, 2d, and the quasi-isothermal halo profile.
%The most optimistic profile is model 2f, which is very close to that of
%Berezinsky et al (1992).  Models 4d and 2d are both NFW halo profiles.}
%\label{jpsi-nim.fig}
%\end{center}
%%\end{wrapfigure}
%\end{figure}

The remaining two parameters that determine the \gray\ flux are, from Eq. \ref{dfdo2.eq},
the neutralino mass $M_{\chi}$ and the annihilation cross sections $\sigma_{\chi\chi}$
that result in \gray\ production.  Monochromatic photons from $\chi\chi\rightarrow\gamma\gamma$
and $\chi\chi\rightarrow Z^{0}\gamma$ are necessarily produced via box diagrams \cite{ull97,ber97a},
and so have a smaller branching ratio that tree-level annihilation into $W$ and $Z^{0}$ pairs.
Secondary \grays\ from the decay of tree-level $W$'s and $Z^{0}$'s give a much larger continuum flux,
which, depending on the SUSY parameters, may be well above the extragalactic \gray\ background
flux (Section \ref{egrb.sec}) near the Galactic center \cite{ber97}.  

\subsubsection{AMS Sensitivity to Monochromatic Gamma Rays from Neutralino Annihilation}
\label{monochromatic.sec}

Monochromatic \gray\ detection is limited both by detector energy resolution (Fig. \ref{energyres.fig})
and aperture.  Should the expected event rate be significantly greater than unity,
the line width must be small enough so that the signal is not lost in the galactic and extragalactic
\gray\ backgrounds.  The rapid deterioration of energy resolution above 100 GeV limits AMS detection 
of {\it monochromatic} \grays\ to neutralino
masses $M_{\chi}<200$ GeV.  (This limit does not apply to detection of continuum photons
from $W$ and $Z^{0}$ decays; see Section \ref{continuum.sec}.)  However, recent analysis by
the DAMA/NaI Collaboration has confirmed the presence of an annual modulation in their detector's
event rate (at 98.5\% C.L.), suggesting the existence of a local WIMP density with a
$1\sigma$ mass range $M_{\chi}=59^{+22}_{-14}$ GeV, and
WIMP-nucleon scalar elastic cross section $\xi\sigma_{\rm scalar}^{\rm (nucleon)}
=7.0^{+0.4}_{-1.7}\times 10^{-9}$ nb, where $\xi$ is the fraction of the local dark matter
density that is due to WIMP neutralinos \cite{brn98}.  Within this mass range our rms energy
resolution is flat at 24\%.
These mass and cross section ranges have also
been shown to be fully compatible with a MSSM neutralino being the major component of 
dark matter in the universe \cite{bot98}.  

\begin{figure}[h]
\begin{center}
\mbox{\epsfig{file=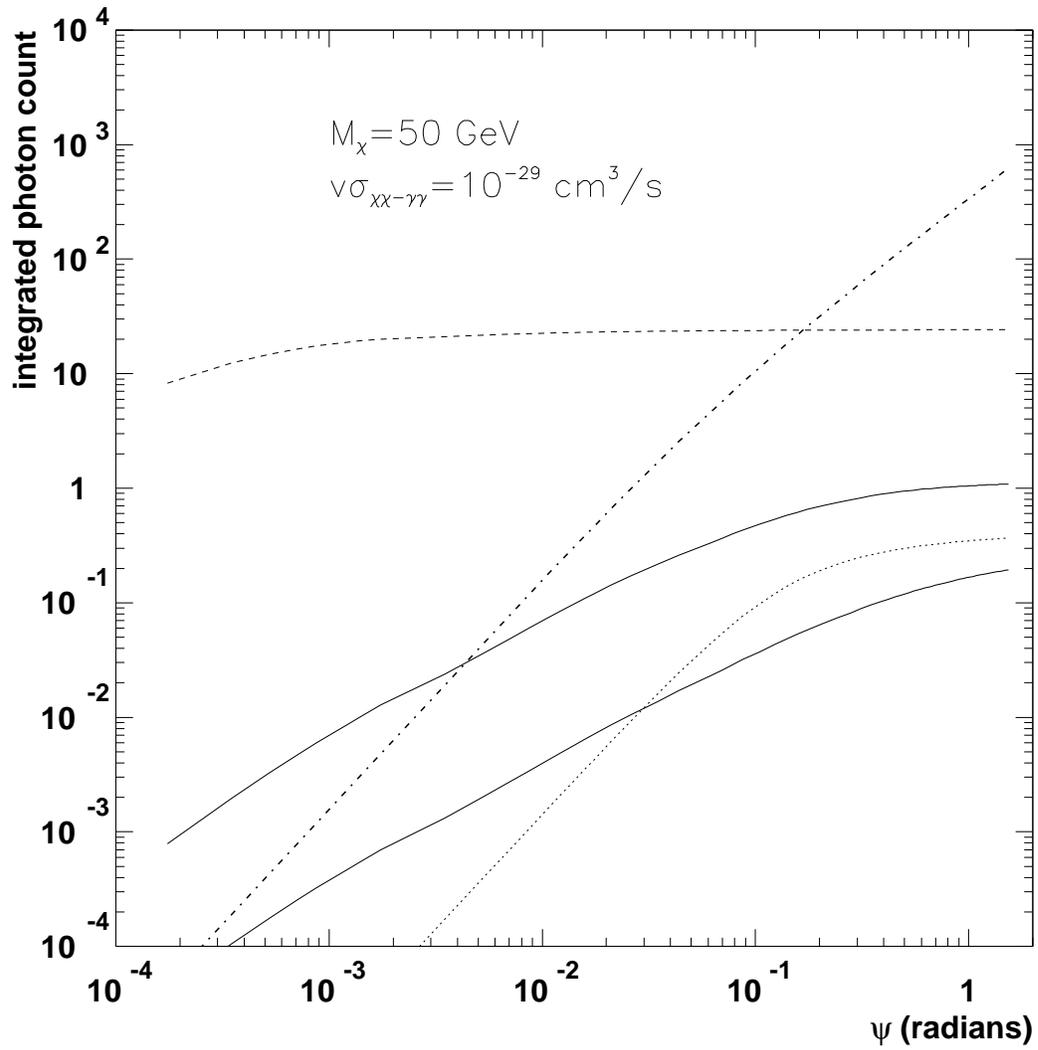,width=6.0in}}
\caption{\small Integrated \gray\ counts $N(<\psi)$ as a function of angle $\psi$ (in radians)
from the galactic center, assuming $M_{\chi}=50$ GeV, producing a narrow line at
$E_{\gamma}=50$ GeV.  See discussion in text.}
\label{mono-nim.fig}
\end{center}
\end{figure}

Figure \ref{mono-nim.fig} 
illustrates the sensitivity of AMS to detection of line photons
from the annihilation channel $\chi\chi\rightarrow \gamma\gamma$ as a function of halo model.  
Plotted are integrated \gray\ counts $N(<\psi)$ as a function of
angle $\psi$ from the Galactic center for the four halo profile models, accumulated over a 
two-year period by AMS, assuming $M_{\chi}=50$ GeV and a velocity-cross section product
$v\sigma=10^{-29}$ cm$^{3}$/s.
With an energy acceptance
width of 20 GeV, approximately 80\% of the line photons are captured.
The dot-dashed line shows the galactic and extragalactic background count for this energy
window; the parameterization of Bergstr\"{o}m, Ullio, and Buckley \cite{ber97}
was used for the galactic
background contribution.  The dashed and dotted lines correspond respectively to
the ``Berezinsky-like'' halo (model 2f of Dehnen and Binney \cite{deh96}) 
and the quasi-isothermal halo, while the upper and lower
solid lines correspond to models 4d and 2d of Dehnen and Binney, which are the
NFW-like models.

%\begin{figure}
%\begin{center}
%\mbox{\epsfig{file=mono99-nim.eps,width=8cm}}
%\caption{\small Integrated \gray\ counts $N(<\psi)$ as a function of angle $\psi$ (in radians)
%from the galactic center, assuming $M_{\chi}=50$ GeV, producing a narrow line at
%$E_{\gamma}=50$ GeV.  See discussion in text.}
%\label{mono-nim.fig}
%\end{center}
%\end{figure}

\begin{figure}[h]
\begin{center}
\mbox{\epsfig{file=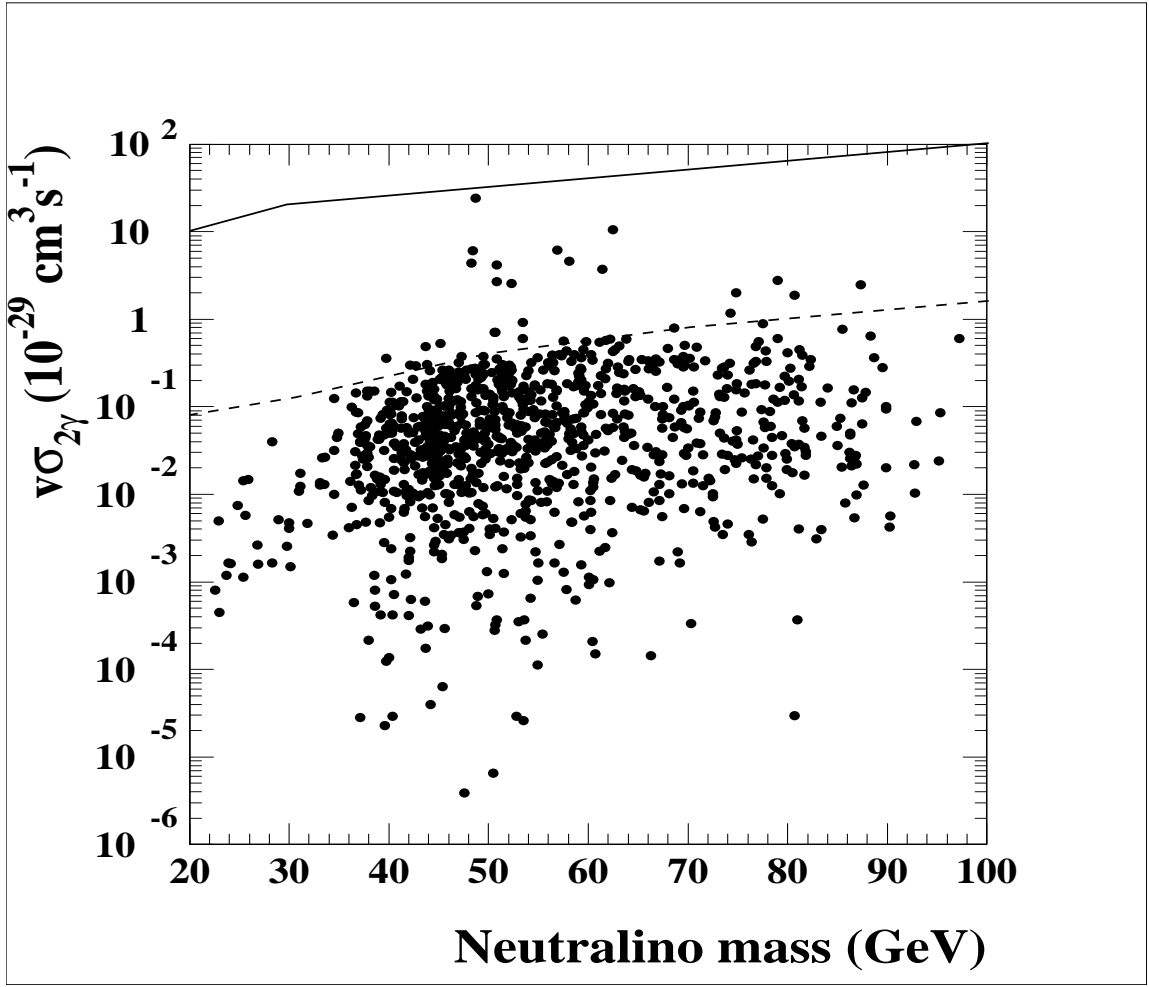,width=6.0in}}
\caption{\small The velocity-cross section product required for AMS detection 
of a neutralino induced \gray\ line as a function
of neutralino mass and halo profile model.  The dashed line corresponds to D\&B model
2f, which is the ``Berezinsky-like'' model; the solid line corresponds to D\&B model
4d, the more productive of the two ``NFW-like'' halo profiles.  These curves are superimposed
on Fig. 3a of Bergstr\"{o}m and Ullio (1997), which gives their calculated cross sections
based on a scan of SUSY parameter space. }
\label{twogamma-nim.fig}
\end{center}
\end{figure}

Bergstr\"{o}m and Ullio \cite{ber97a}
have recently performed a full one-loop calculation of the
$\chi\chi\rightarrow\gamma\gamma$ annihilation reaction, assuming the minimal supersymmetric
extension of the standard model (MSSM).  They performed extensive scans in SUSY parameter
space to map out the range of values of the annihilation cross section versus neutralino mass.
We have reproduced their Figure 3a in Figure \ref{twogamma-nim.fig}, along with the AMS
sensitivity for the two most optimistic halo profiles, models 2f and 4d, expressed as a lower
limit on $\left<v\sigma\right>$ versus $M_{\chi}$ for detection of neutralino annihilation.
The detection criteria were arbitrarily chosen to be $N_{\gamma}\ge 10$ and
$N_{\gamma}/\sqrt{B_{\gamma}}\ge 6$, where $N_{\gamma}$ is the total number of annihilation
\grays\ detected over a 2-year operating period within the central few degrees of the 
Galactic Center, and $B_{\gamma}$ is the total number of background photons from diffuse
galactic emission detected within the energy window and the same solid angle about the
Galactic Center.

%\begin{figure}
%\begin{center}
%\mbox{\epsfig{file=twogamma99-nim.eps,width=8.5cm}}
%\caption{\small The velocity-cross section product required for AMS detection 
%of a neutralino induced \gray\ line as a function
%of neutralino mass and halo profile model.  The dashed line corresponds to D\&B model
%2f, which is the ``Berezinsky-like'' model; the solid line corresponds to D\&B model
%4d, the more productive of the two ``NFW-like'' halo profiles.  These curves are superimposed
%on Fig. 3a of Bergstr\"{o}m and Ullio (1997), which gives their calculated cross sections
%based on a scan of SUSY parameter space. }
%\label{twogamma-nim.fig}
%\end{center}
%\end{figure}

\begin{figure}[h]
\begin{center}
\mbox{\epsfig{file=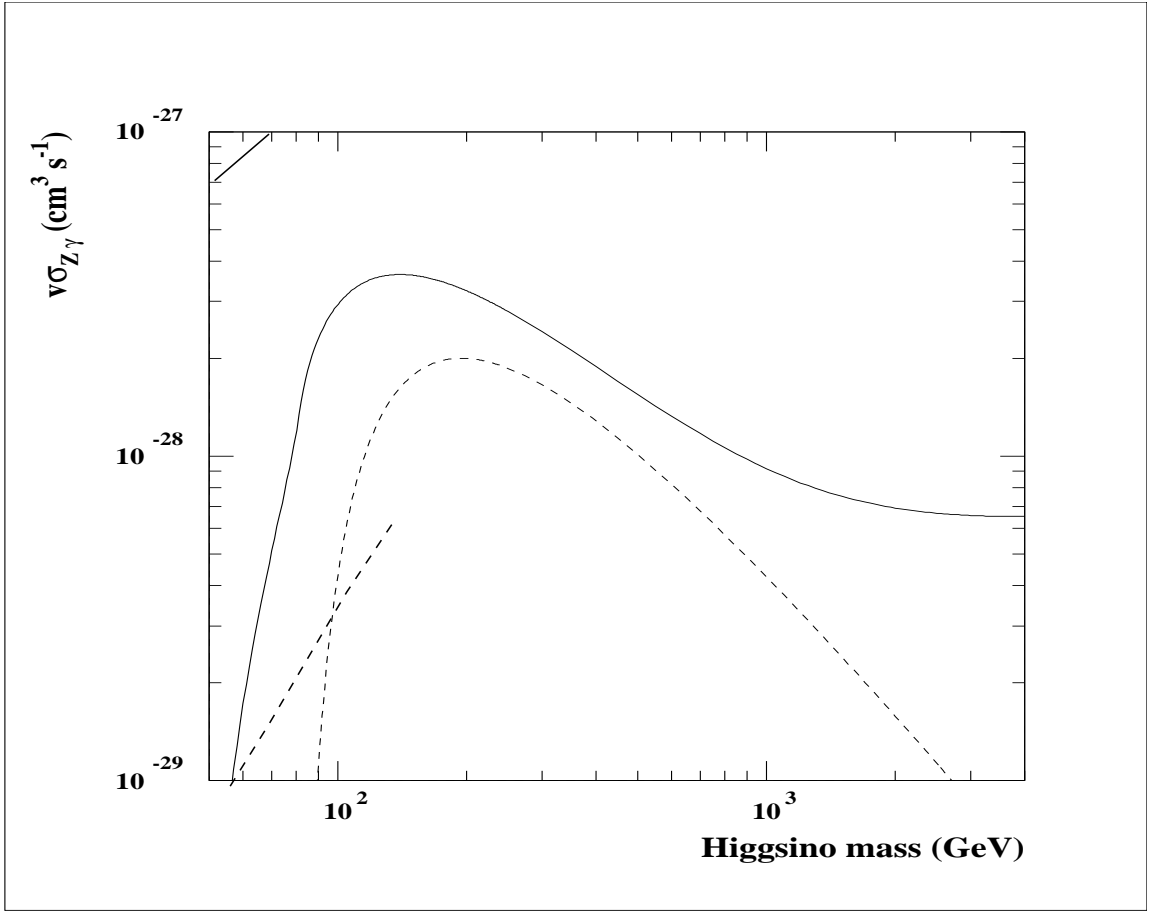,width=6.0in}}
\caption{\small The velocity-cross section product required for AMS detection as a function
of neutralino mass and halo profile model. 
These curves are from
Fig. 6 of Ullio and Bergstr\"{o}m (1997), which gives their calculated cross section
for the single-photon production channel $\chi\chi\rightarrow\gamma Z$ 
or a pure higgsino state.  The dashed-line line segment corresponds to D\&B model
2f, which is the ``Berezinsky-like'' model.  The other halo profiles do not intersect
the Ullio and Bergstr\"{o}m curves.  (The solid line segment in the upper-left corner
corresponds to D\&B model 4d.)}
\label{onegamma-nim.fig}
\end{center}
\end{figure}

Ullio and Bergstr\"{o}m \cite{ull97} also performed a MSSM full one-loop calculation of 
neutralino annihilation via $\chi\chi\rightarrow\gamma Z$, assuming a pure higgsino
state for the neutralino.  For this channel the \gray\ has an energy
$E_{\gamma}=M_{\chi}-m_{Z}^{2}/4M_{\chi}$.  In
Figure \ref{onegamma-nim.fig} we have reproduced Figure 6 of Ullio and Bergstr\"{o}m,
which shows their $v\sigma_{Z\gamma}$ versus higgsino mass (solid line), along with
the unitary lower bound on this cross section (dashed line).  Superimposed on this
plot is a line segment which shows the minimum  $v\sigma_{Z\gamma}$ required for
detectability by AMS over a 2-year period, assuming
the ``Berezinsky-like'' halo model 2f of Dehnen and Binney \cite{deh96}.  None of the other
halo models intersect the Ullio and Bergstr\"{o}m curves.  Thus
for this pure higgsino 
annihilation channel only the most
singular halo profile can yield a detectable signal.

%\begin{figure}
%\begin{center}
%\mbox{\epsfig{file=onegamma99-nim.eps,width=8.5cm}}
%\caption{\small The velocity-cross section product required for AMS detection as a function
%of neutralino mass and halo profile model. 
%These curves are from
%Fig. 6 of Ullio and Bergstr\"{o}m (1997), which gives their calculated cross section
%for the single-photon production channel $\chi\chi\rightarrow\gamma Z$ 
%for a pure higgsino state.  The straight solid line segment corresponds to D\&B model
%2f, which is the ``Berezinsky-like'' model; the other halo profiles do not intersect
%the Ullio and Bergstr\"{o}m curves.}
%\label{onegamma-nim.fig}
%\end{center}
%\end{figure}

In general, then, only the more singular of the halo profile models, in
which the dark matter density continues to increase as $R^{-\gamma}$,
$\gamma\sim 1-2$, can give hope for line photon detection by AMS, 
and then only in a restricted region of MSSM parameter space.  It should be
kept in mind that a fraction of this restricted region may already have been
eliminated by the absence of a detection by EGRET, but such an analysis has yet
to be completed \cite{din98}.

\subsubsection{AMS Sensitivity to Continuum Gamma Rays from Neutralino Annihilation}
\label{continuum.sec}

Continuum photons from the decay of $W$s and $Z$s produced by tree-level annihilations
$\chi\chi\rightarrow WW,ZZ$ are far more abundant than the line \grays\ considered above,
and so possibly provide the greatest chance for detection, assuming the neutralino mass
is above that of the $W$ and $Z$.  Berstr\"{o}m, Ullio, and Buckley \cite{ber97} have used the
PYTHIA Monte Carlo code \cite{sjo94} to determine the 
the continuum photon energy distribution from $\chi\chi$ annihilation into $WW$, $ZZ$,
and $q\bar{q}$ pairs.  They find that the hardest distributions come from $WW$ and $ZZ$
pairs, and that the annihilation rate is dominated by $WW$ and $ZZ$ production at
higher (higgsino-like) neutralino masses.  From a scan of MSSM parameter space, they
find that the models with the highest rate of $WW$ and $ZZ$ production yield an
observed differential photon flux of 
\begin{equation}
\frac{dF_{\gamma}}{dE_{\gamma}}\approx 1.2\times 10^{-10}
\left(\frac{300 {\rm \ GeV}}{M_{\chi}}\right)^{4}\frac{dN_{\gamma}}{dE_{\gamma}}
J(\psi) {\rm \ \ cm}^{-2}{\rm s}^{-1}{\rm sr}^{-1}{\rm GeV}^{-1},
\end{equation}
where
\begin{equation}
\frac{dN_{\gamma}}{dx}=M_{\chi}\frac{dN_{\gamma}}{dE_{\gamma}}=
\frac{0.73}{x^{1.5}}e^{-7.8x},
\end{equation}
for $M_{\chi}>300$ GeV. The $J(\psi)$ factor is given by Eq. \ref{jpsi.eq}.

\begin{figure}[h]
\begin{center}
\mbox{\epsfig{file=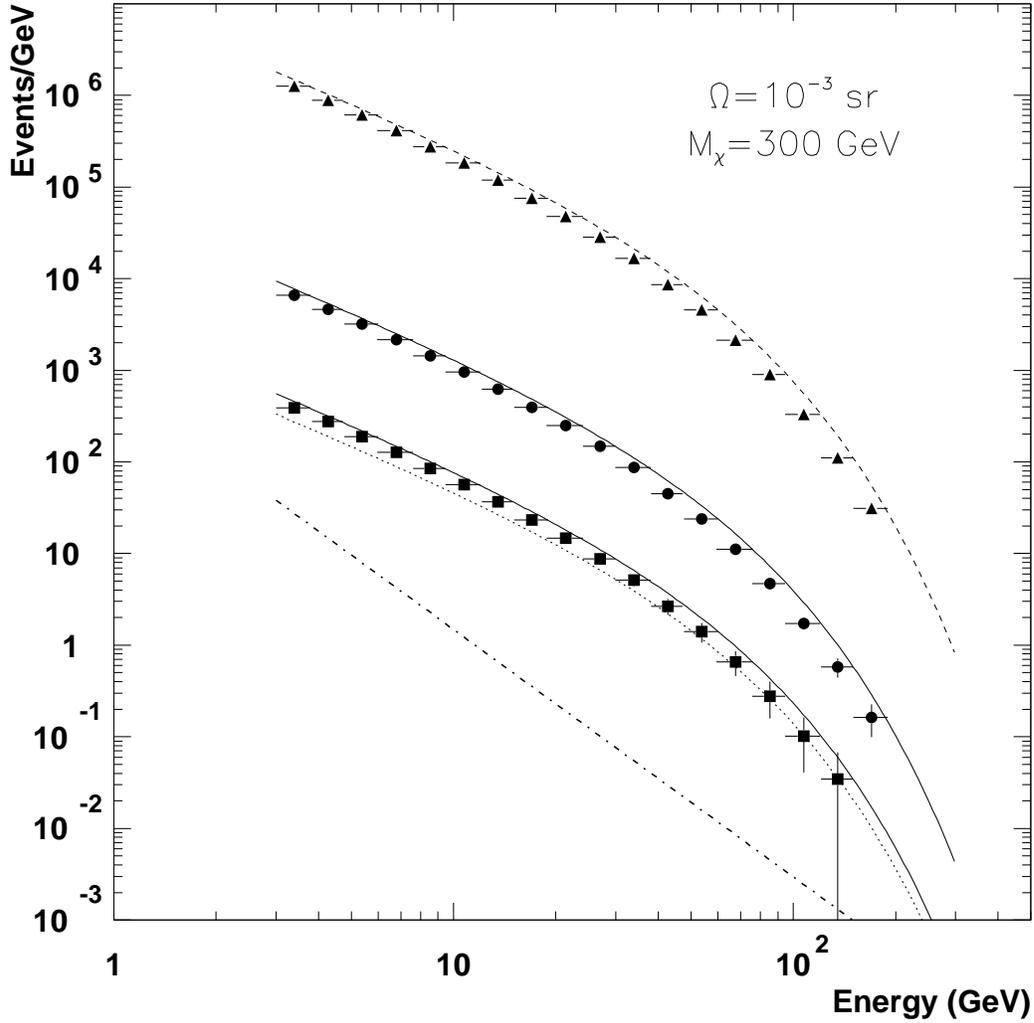,width=6.0in}}
\caption{\small The differential continuum \gray\ event count
from neutralino (higgsino)
annihilation in the Galactic halo detected by AMS over a 2-year time period,
within a solid angle of $10^{-3}$ sr about the Galactic center, based on
the cross sections for $\chi\chi\rightarrow WW,ZZ$ of Bergstr\"{o}m, Ullio,
and Buckley \cite{ber97}.  The dashed, upper and lower solid, and dotted lines correspond
to the various halo models:  Models 2f, 4d, 2d, and quasi-isothermal, respectively.
The dot-dashed line
shows the sum of the integrated diffuse galactic and extragalactic \gray\ background
\grays\ detected by within the same solid angle.  The faux-data points show
the energy bin counts obtained by AMS after convolution with the energy resolution
function of the detector.  (Faux data points are not shown for the quasi-isothermal
halo model.)}
\label{cont.300.1-nim.fig}
\end{center}
\end{figure}

\begin{figure}[h]
\begin{center}
\mbox{\epsfig{file=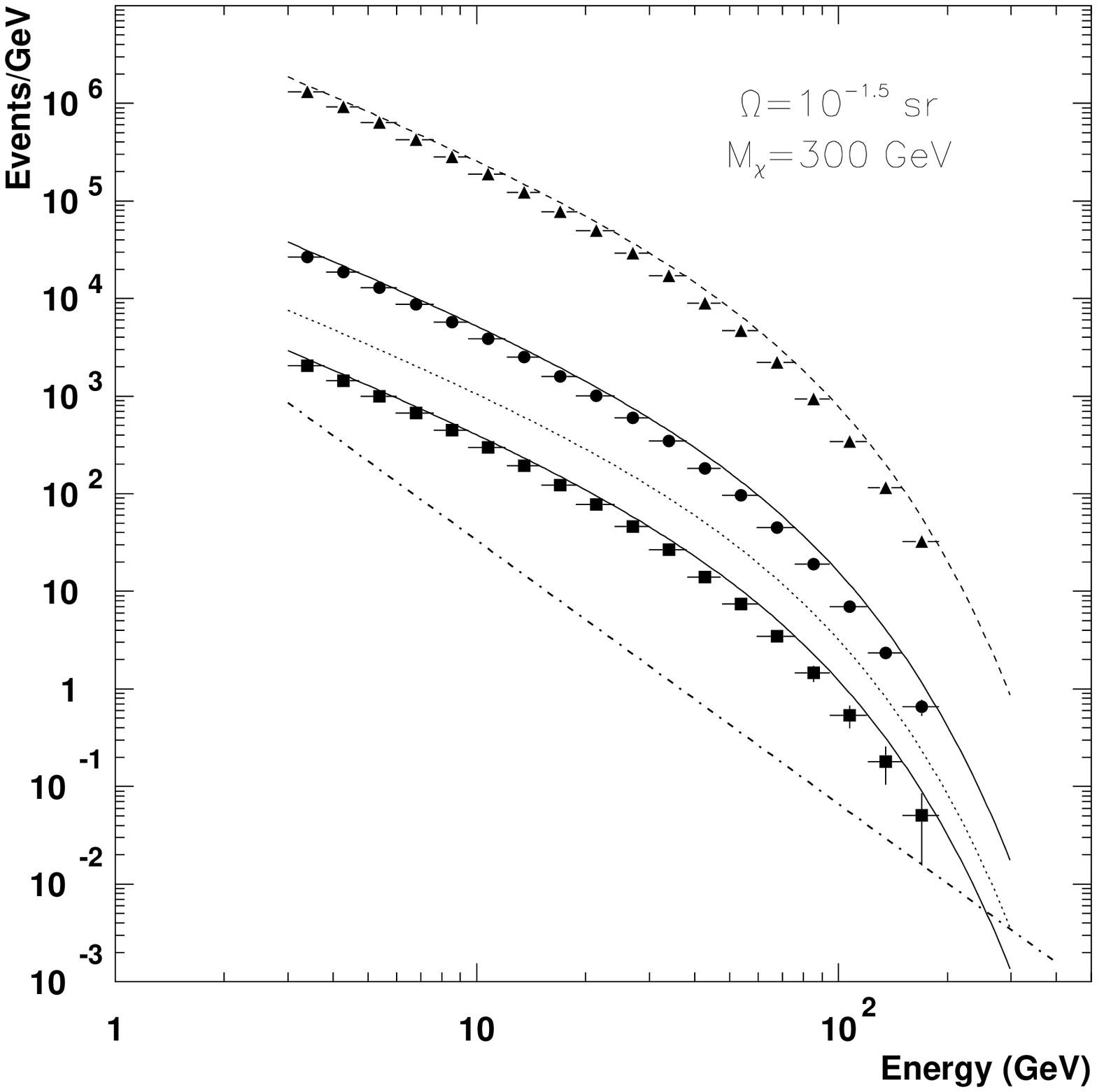,width=6.0in}}
\caption{\small The same as Figure \ref{cont.300.1-nim.fig}, except the fluxes are
those that arrive within
a solid angle of 0.03 sr about the Galatic center.  }
\label{cont.300.6-nim.fig}
\end{center}
\end{figure}

\begin{figure}[h]
\begin{center}
\mbox{\epsfig{file=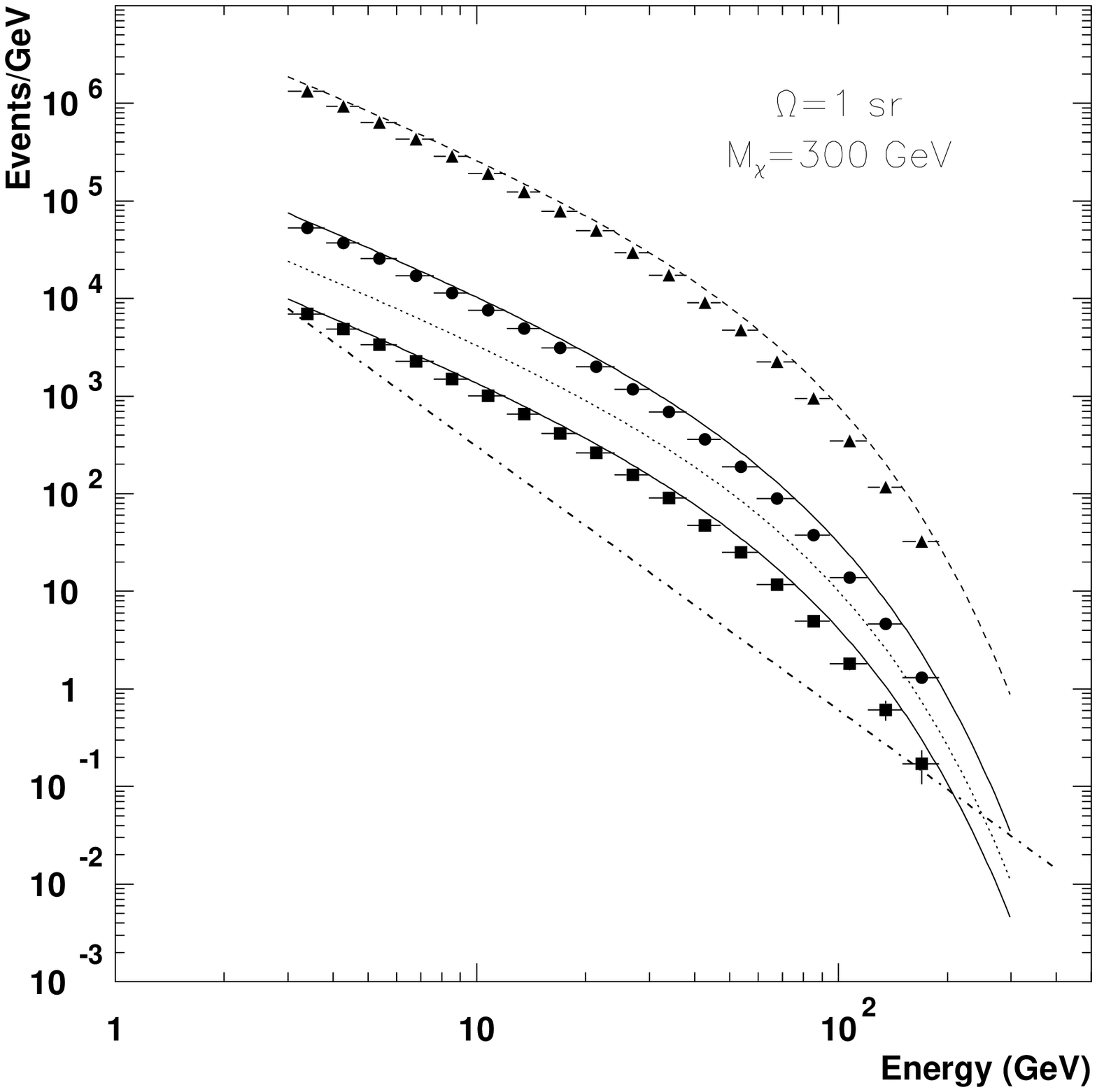,width=6.0in}}
\caption{\small The same as Figure \ref{cont.300.1-nim.fig}, except the fluxes are
those that arrive within
a solid angle of 1 sr about the Galatic center.  }
\label{cont.300.32-nim.fig}
\end{center}
\end{figure}

Using the $J(\psi)$ functions of the four halo models described above 
(Fig. \ref{jpsi-nim.fig}), the differential flux coming from a solid angle
$\Omega$ about the Galactic Center can be obtained and compared to the flux
from the galactic and extragalactic backgrounds.  Using the collection area*time
product $At(E_{\gamma},\delta)$ (Fig. \ref{areatime.fig}) for AMS, the differential
event count is obtained.  Figures \ref{cont.300.1-nim.fig} to
\ref{cont.300.32-nim.fig} show these event counts after integration over $\psi$
out to three different solid angles
about the Galactic Center, $\Omega=10^{-3}$, $10^{-1.5}$, and 1.0 steradians.
The dot-dashed line is the diffuse galactic and extragalactic background photon
count; the upper and lower solid lines correspond to halos for models 4d and 2d,
respectively; the dashed line corresponds to the ``Berezinsky-like'' halo model
2f; and the dotted line comes from the quasi-isothermal halo model.  The faux-data
points shown in the figures take into account the effect of detector energy resolution; these
are the counts obtained in each energy bin after convolution with the Monte Carlo
energy resolution functions (Fig. \ref{energy.fig}).

There are a number of salient features, the primary one being the strength of the
signal compared to the diffuse background for essentially all of the halo models
considered.  For this particular (optimized) MSSM model it is likely that EGRET would have
already provided a detection given any of the halo models considered here.
Although there is a large variation in signal between the various halo models, it appears
nevertheless that for {\it any} of these halo models
a substantial region of MSSM parameter space can be constrained
by the absence of a continuum \gray\ signal.

These figures also show that the differing angular dependences of the background and
signal result in there being an optimal solid angle $\Omega$ that maximizes signal
to noise, with more singular profiles having a stronger signal to noise at smaller
solid angles.  These differing angular dependences can also be used to discriminate
the signal from the diffuse backgrounds.

Finally, we point out that
continuum photons result from $\chi\chi\rightarrow b\bar{b}$ as well as from
$WW$ and $ZZ$ pair production, producing scaled spectra similar to those from
$W$ and $Z$ decays.  This implies that \amsg\ sensitivity to continuum \grays\ 
may extend down to the current LEP limits on the neutralino mass \cite{bia98}.

\section{Conclusions}
\label{conclusions.sec}
We have shown that with minor modifications the Alpha Magnetic Spectrometer can
become a powerful \gray\ detector as well, with overall performance characteristics
being comparable, if not superior, to those of EGRET.  With \gray\ energy
resolution extending past 100 GeV, and with an aperture that is nearly 
flat above $\sim 3$ GeV, \amsg\ can address a number of outstanding issues in 
\gray\ astrophysics that relate to the relatively unexplored region of
$E_{\gamma}=20-200$ GeV.  For one, \amsg\ will likely confirm or refute
the hypothesis that unresolved blazars are responsible for the bulk of the
extragalactic \gray\ background; \amsg\ will also extend the spectrum of the
diffuse galactic background to above 100 GeV, helping to resolve current
difficulties in interpreting the EGRET diffuse galactic background 
measurement \cite{poh98}.

\amsg\ should roughly double the total number of blazars detected in \grays, and
will be enable multiwavelength observational campaigns to include the GeV region
of blazar spectra during the flight years of 2003-2006.  There is an additional
possibility that an indirect detection of the cosmic UV and optical photon
background can be made through the detection of extinctions in high-redshift
blazars above $\sim 20$ GeV.  \amsg\ will also likely observe GeV \gray\ emission
from one or more gamma ray bursts during its operational lifetime.

\amsg\ will also search for both line and continuum emission of \grays\ from the
region of the Galactic Center created by the annihilation of dark matter neutralinos.
Although the sensitivity to line emission appears marginal, there is nevertheless 
a finite, though small, region of halo/MSSM phase space which allows a detection by AMS.
However, a much larger region of dark-matter halo/MSSM parameter space can be constrained in
a search for continuum \grays\ by \amsg.

Finally, we note that even higher sensitivities can be reached with the addition
of a high granularity calorimeter below the magnet (option (b)); this will be the
subject of future work.

\vspace{3ex}

\begin{center}
{\bf Acknowledgements}
\end{center}
We thank B. Dingus, S. Hunter, and P. Ullio for helpful conversations, and P. Sreekumar for
the same and for supplying us with additional, unpublished information on 
the EGRET measurement of the extragalactic \gray\ background.  We also thank the many
members of the AMS Collaboration who have indirectly contributed to this paper, and
especially S.C.C. Ting for his critical comments.

\newpage

\setcounter{totalnumber}{1}

\newpage

\newpage

\newpage

\newpage

\newpage

\newpage

\newpage

\newpage

\newpage

\newpage

\newpage

\newpage

\newpage

\newpage

\newpage

\newpage

\newpage

\newpage

\newpage

\end{document}